\documentclass{an}
\usepackage{graphicx}
\usepackage{times}
\Volume{01}              
\Year{2005}              
\Month{01}               
\Pagespan{1}{7}      	 

\begin{document}
\lhead[\thepage]{A.N. Bilir et al.: Absolute magnitudes for late-type dwarf stars for Sloan photometry}
\rhead[Astron. Nachr./AN~{\bf XXX} (200X) X]{\thepage}
\headnote{Astron. Nachr./AN {\bf 32X} (200X) X, XXX--XXX}

\title{Absolute magnitudes for late-type dwarf stars for Sloan photometry}

\author{S. Bilir   
\and  S. Karaali
\and  S. Tun\c{c}el}
\institute{Istanbul University Science Faculty, 
           Department of Astronomy and Space Sciences, 
           34119, University-Istanbul, Turkey}
\date{} 

\abstract{We present a new formula for absolute magnitude determination for 
late-type dwarf stars as a function of $(g-r)$ and $(r-i)$ for $Sloan$ 
photometry. The absolute magnitudes estimated by this approach are 
brighter than those estimated by colour-magnitude diagrams, and they reduce 
the luminosity function rather close to the luminosity function of Hipparcos.
\keywords{Techniques: Sloan photometry -- Stars: distances -- 
Stars: luminosity function}
}
\correspondence{sbilir@istanbul.edu.tr}

\maketitle

\section{Introduction}
Absolute magnitudes play an important role in the Galactic research due to 
the opportunity for distance determination by its combination with the apparent 
magnitude of the star considered. This method is used especially for distance 
stars for which trigonometric parallaxes are not available. However, the 
procedures may be different. One of them is to use a specific colour-magnitude 
diagram, where the works of Phleps et al. (2000), Chen et al. (2001), 
Siegel et al. (2002), Karaali et al. (2003a), Du et al. (2003), Bilir, Karaali 
\& Buser (2004), Karaali, Bilir \& Buser (2004a), Karata\c{s} et al. (2004), 
and Karaali, Bilir \& Hamzao\u glu (2004b) can be given as examples. This 
procedure needs separation of stars into different metallicity intervals and 
choosing the appropriate colour-magnitude diagram. For example, the colour 
magnitude diagrams of the clusters M92 and 47 Tuc are usually used 
for halo and thick disk populations, respectively, and the colour-magnitude 
diagram given by Lang (1992) for thin disk population (cf. Karaali et al. 2004b).

An alternative approach for absolute magnitude determination is to use the 
absolute magnitude offset from a specific colour-magnitude diagram. The offset 
from the Hyades cluster given by Laird et al. (1988) is a function of both 
$UBV$ colour index and $UV$-excess which  measures the metallicity. Karaali 
et al. (2003b) and Karaali, Bilir \& Tun\c{c}el (2005) calibrated the absolute 
magnitude offset with the $UV$-excess for different $UBV$ colour-indices and 
they extended their work from $UBV$ to $Sloan$ photometry. Thus they improved 
this procedure considerably. However the work carried out by these authors is 
limited with $0.3<(B-V)\leq1.1$ for $UBV$ or $0.09<(g-r)\leq0.93$ for 
$Sloan$ photometry (Karaali, Bilir \& Tun\c{c}el 2005). The colour index 
$(g-r)=0.93$ is close to the limiting value of $(g-r)=1.1$ 
which separates thin and thick disk and halo couple for $Vega$ photometry, 
as claimed by Karaali et al. (2004b) (we should remind that the bands for 
$g$ and $r$ are almost the same for $Sloan$ and $Vega$ photometry. 

Karaali et al. (2003b) showed that a single colour-magnitude diagram does not 
supply reliable absolute magnitudes for stars with a large range of 
metallicities, such as halo for example. Hence, it seems that absolute magnitude 
estimation by means of offset from the colour-magnitude diagram of Hyades, as 
a function of colour-index, has advantages on the other procedure for thick 
disk and halo populations. However, for thin disk where the metallicity range is 
not large, one can use a colour-magnitude diagram with $[Fe/H]\sim0$ metallicity. 

Although the transformation equations of Smith et al. (2002) enable $M(g^{'})$ 
absolute magnitude estimation for $Sloan$ photometry, $M(g^{'})$ is only 
$(g^{'}-r^{'})$ dependent in his formulae. As $(r-i)$ is insensitive to 
$(g-r)$ for late-type stars, we though to derive an equation for the 
absolute magnitude $M(g)$ which covers both $(g-r)$ and $(r-i)$, and this is 
the main topic of this paper. We will see that this approach provides 
more appropriate absolute magnitudes.         	       

\begin{figure*}
\center
\resizebox{16cm}{6.40cm}{\includegraphics*{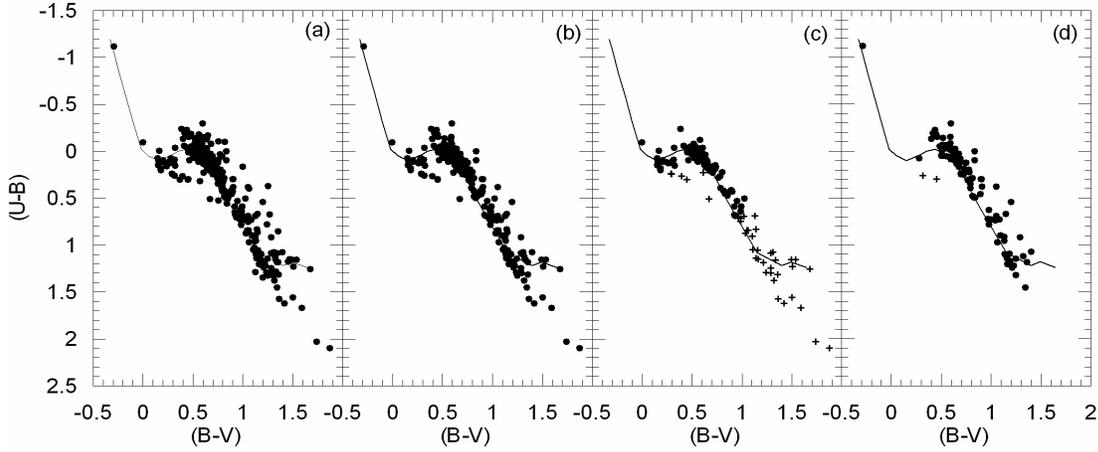}} 
\caption {Four $(U-B, B-V)$ two-colour diagrams for stars taken or supplied 
from different sources (see text): (a) for all stars (307), (b) for 237 stars 
with colour or magnitude errors less than $0^{m}.05$, (c) and (d) the same as 
in (b) but for different apparent magnitudes. For (c) $V<13^{m}$ and (d) 
$V>13^{m}$. The symbol ($+$) shows the positions of giants.}
\end {figure*}

\begin{figure*}
\center
\resizebox{15cm}{4.65cm}{\includegraphics*{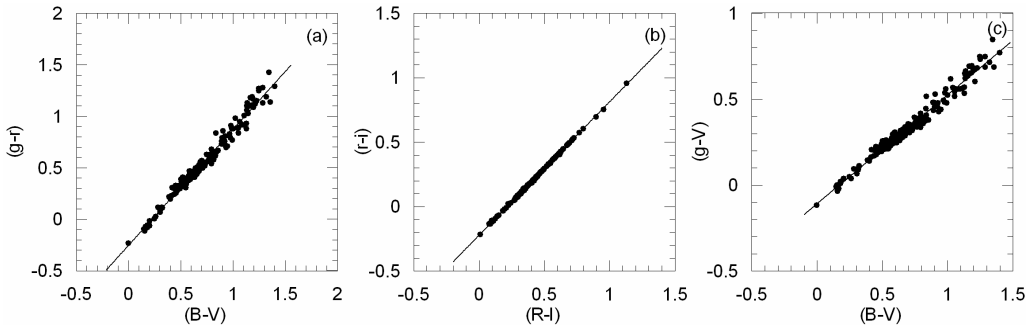}} 
\caption {Three different two-colour diagrams for the dwarf sample: (a) 
$(g-r, B-V)$, (b) $(r-i, R-I)$, and (c) $(g-V, B-V)$.}
\end {figure*}

\section{The data and revision of the transformation formulae of Smith et al. (2002)}

The $UBVR_{C}I_{C}$ and $ugriz$ magnitudes are available for 307 stars. The $ugriz$ 
data for 79 stars are transformed from the $u^{'}g^{'}r^{'}i^{'}z^{'}$ data of 
Smith et al. (2002) by means of the transformation formulae of Rider et al. (2004), 
whereas those of 228 stars which are standards of Landolt (1992)  are supplied from 
Cambridge Astronomical Institute (Derek Jones, private communication). The data are 
given in Table 1. The $(U-B, B-V)$ two-colour diagram for 307 stars shows large 
scattering at the location where metal-poor stars are dominant. Also, a long and 
narrow tail extends to large $(U-B)$ and $(B-V)$ colour indices, indicating to 
the existence of late-type giants (Fig. 1a). The scattering reduced when stars with 
errors in two-colours larger than $0^{m}.05$ are omitted (Fig. 1b), and 34 stars 
at the position of the tail plus eight stars below the main-sequence with 
$0.2<(U-B)<0.5$ were adopted as giants (Fig. 1c-d). Thus the sample reduced to 
195 dwarf stars. 

We plotted the stars of the sample in the two-colour diagrams $(g-r, B-V)$ 
and $(r-i, R-I)$, and we derived the following transformation equations 
(Fig. 2a and Fig. 2b, respectively):

\begin{equation}
(g-r) = 1.12431(B-V)-0.25187\\
\end{equation}

\begin{equation}
(r-i) = 1.04036(R-I)-0.22359\\
\end{equation}

The third transformation equation concerns with the absolute magnitude. 
The two-colour diagram for $(g-V, B-V)$ is given in Fig. 2c for 
which the following transformation equation could be derived:

\begin{equation}
g-V = 0.63359(B-V)-0.10813\\
\end{equation}

From equation (3) one can obtain the transformation equation for the absolute 
magnitude $M(g)$ for $Sloan$ photometry:

\begin{equation}
M(g)=M(V)+0.63359(B-V)-0.10813\\
\end{equation}

We used the $UBVR_{C}I_{C}$ data for 708 solar neighbourhood stars with 
relative parallax errors less than 0.1, from the WEB page of Neill Reid 
\footnote{http://www-int.stsci.edu/$\sim$inr/cmd.html} to test the 
transformation formulae (1) and (2), and to compare them with the ones of 
Smith et al. (2002). Fig. 3a shows the $(g-r, r-i)$ two-colour diagram for 
the data transformed by the mentioned equations. The locus points shown in 
the same diagram belong to the bins evaluated for 6049  stars in a field 
in the anti-centre direction of the Galaxy ($l=181^{o}.88$, $b=-45^{o}.19$, 
epoch 2000, size 8 square-degrees), taken from DR1 \footnote{http://www.sdss.org/dr1/}. 
The good agreement between the locus points and the positions of 708 stars 
mentioned above, for $(g-r)<1.4$, confirm the transformation equations (1) 
and (2). Whereas, the positions of 708 stars with $(g-r)$ and $(r-i)$ colour 
indices estimated by means of the combined transformation formulae of Smith 
et al. (2002) and Rider et al. (2004) deviate from the locus points for 
$(g-r)>0.9$ considerably (Fig. 3b).	 				

\begin{figure}
\center
\resizebox{8cm}{14.80cm}{\includegraphics*{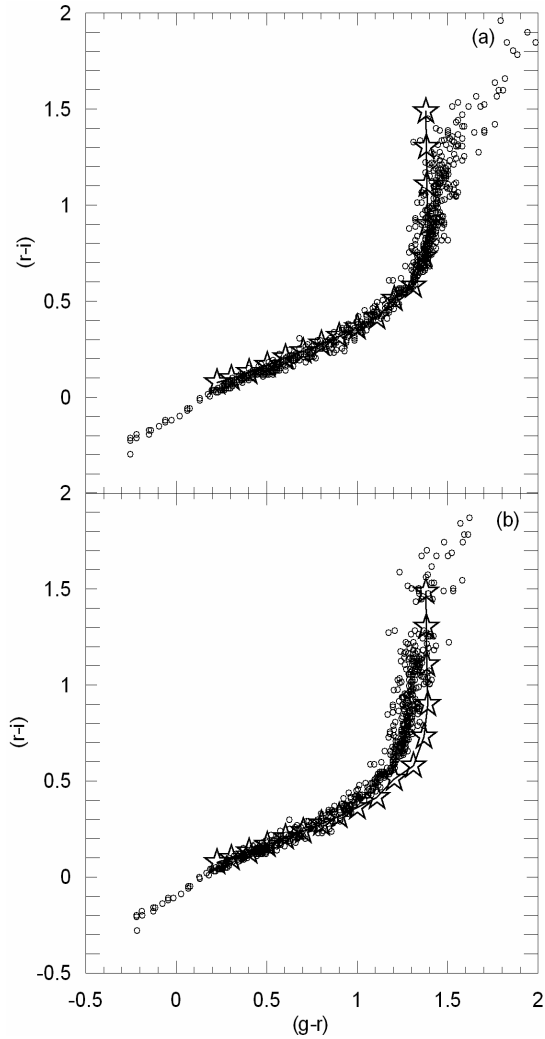}} 
\caption {$(g-r, r-i)$ two-colour diagram for colours 
transformed from $UBVR_{C}I_{C}$ data of Neill Reid for 708 stars by different 
formulae, confronted to the colours of 6049 stars in a field in the 
anti-centre of the Galaxy: (a) by the new formulae, (1) and (2); (b) by 
modified the formulae of Smith et al. (2002). The large asterisks show the 
locus points of 6049 stars. There is an agreement between two star samples 
in the panel (a) for $(g-r)<1.4$, whereas the trends for two 
samples are different for $(g-r)>0.9$ in panel (b).}
\end {figure}

\section{Absolute magnitude as a function of two-colour indices}

We derived a two-colour dependent equation for the absolute magnitude $M(g)$, 
for $Sloan$ photometry, by means of a series of calculations as follows. Let us 
remind that our purpose in this work is the estimation of $M(g)$ absolute 
magnitudes for late-type stars with $(g-r)>0.93$ magnitudes for which 
the procedure of Karaali et al. (2003b) is not available. First we used the $M(V)$ 
and $(B-V)$ data of Henry et al. (1999) and we transformed the $(B-V)$ colour 
indices to $(R-I)$ by the relation of Cox (2000). The ranges for these data are 
$0.90<(B-V)\leq1.60$, $0.45<(R-I)\leq1.85$, and $6<M(V)\leq11.5$. Second, we used 
these data and the appropriate  equation, i.e. (1), (2) or (4), for estimation 
$(g-r)$, $(r-i)$ and $M(g)$. Then, third and finally, we estimated the 
coefficients $a$, $b$, and $c$ in the following equation by means of the 
least-square method 

\begin{equation}
M(g)=a(g-r)+b(r-i)+c\\
\end{equation}
The numerical values of the coefficients are: $a=5.791$, $b=1.242$, and $c=1.412$.

\section {Testing the absolute magnitudes in luminosity function}

We tested the absolute magnitudes evaluated by equation (5) in the luminosity 
function for stars in the field mentioned in Section 2. We followed the 
following procedure, for this purpose.

\subsection{Density functions}

We have two sets of density functions: For the first set, the absolute magnitudes 
were evaluated by equation (5), whereas for the second set we used the following 
modified equation of Smith et al. (2002) by means of the transformation formulae 
of Rider et al. (2004):

\begin{equation}
M(g)= M(g^{'})+0.059(B-V)-0.043\\
\end{equation}
where $M(g^{'})$ is the absolute magnitude in the system of Smith et al. (2002), i.e.
\begin{equation}
M(g^{'})= M(V)+0.54(B-V)-0.07\\
\end{equation}
The combination of this equation with the equations of Rider et al. (2004) and Smith 
et al. (2002), cited in the following respectively, gives equation (6):   

\begin{equation}
g = g^{'} + 0.06[(g^{'}-r^{'})-0.53]
\end{equation}

\begin{equation}
(g^{'}-r^{'}) = 0.98(B-V)-0.19
\end{equation}
The distance $r$ to a star relative to the Sun is carried out by the well 
known formula, i.e.

\begin{equation}
[g-M(g)]_{0}=5\log(r)-5\\
\end{equation}
and its vertical distance to the Galactic plane ($z$) could be evaluated 
by its distance $r$ and its Galactic latitude ($b$):

\begin{equation}
z=r\sin(b)\\
\end{equation}

\begin{figure}
\center
\resizebox{7cm}{12cm}{\includegraphics*{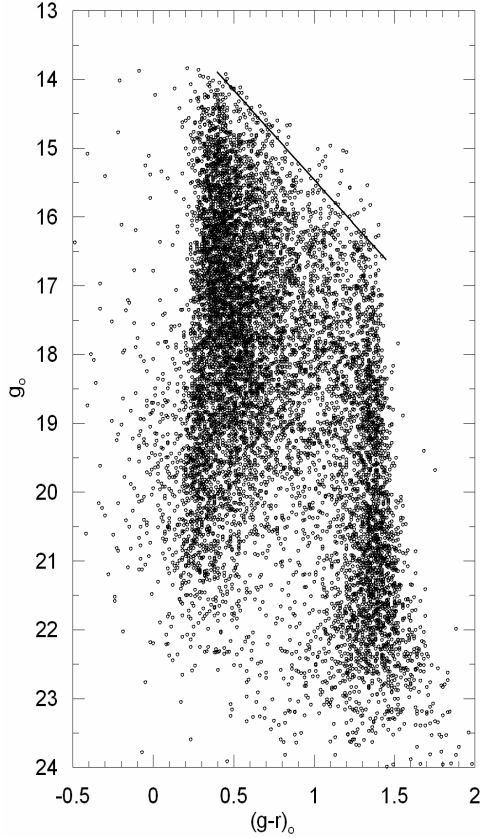}} 
\caption {Colour-magnitude diagram for 6049 stars taken from the field in the 
anti-centre direction of the Galaxy. The bright apparent limiting magnitude of 
these stars is determined by means of the bar at the top of the diagram and the 
limiting $(g-r)_{0}$ colour index for the corresponding absolute magnitude 
interval.}
\end {figure}

Table 2 and Table 3 give the number of stars as a function of distance $r$ 
relative to the Sun, the corresponding mean distance $z^{*}$ from the Galactic 
plane, and absolute magnitude evaluated by equations (5) and (6) respectively. 
It is interesting, there are only few number of stars fainter than $M(g)=11$ 
in Table 2, not available for density evaluation, whereas there are considerable 
number of stars within absolute magnitude intervals $11<M(g)\leq12$ and 
$12<M(g)\leq13$ in Table 3. The reason of this discrepancy is the different 
procedures used for absolute magnitude evaluation. The  horizontal thick lines 
correspond to the limiting distance of completeness ($z_{l}$) evaluated by the 
following equations:

\begin{equation}
[g-M(g)]_{0}=5\log(r_{l})-5\\
\end{equation}

\begin{equation}
z_{l}=r_{l}\sin(b)\\
\end{equation}
where $g_{0}$ is the limiting apparent magnitude, $r_{l}$ is the limiting 
distance of completeness relative to the Sun, and $M(g)$ is the appropriate 
absolute magnitude $M_{1}$ or $M_{2}$ for the absolute magnitude interval 
$M_{1}-M_{2}$ considered. The limiting apparent magnitude for faint stars is 
constant, i.e. $g_{0}$ is adopted as $g_{0}=19.5$ to increase the 
accuracy, whereas for bright stars it depends on the limiting colour-index 
$(g-r)_{0}$ for a specific absolute magnitude interval (Fig. 4). The 
logarithmic space density functions, $D^{*}=\log(D)+10$, evaluated by means of 
these data are omitted from the tables in order to conserve space, but they are 
presented in Fig. 5 and Fig. 6, where: 
$D=N/ \Delta V_{1,2}$, $\Delta V_{1,2}=(\pi/180)^{2}(\sq/3)(r_{2}^{3}-r_{1}^{3})$, 
$\sq$: size of the field (8 square-degrees), $r_{1}$ and $r_{2}$: the limiting 
distance of the volume $\Delta V_{1,2}$ and N: number of stars per unit 
absolute magnitude. 

\begin{figure}
\center
\resizebox{7cm}{7.50cm}{\includegraphics*{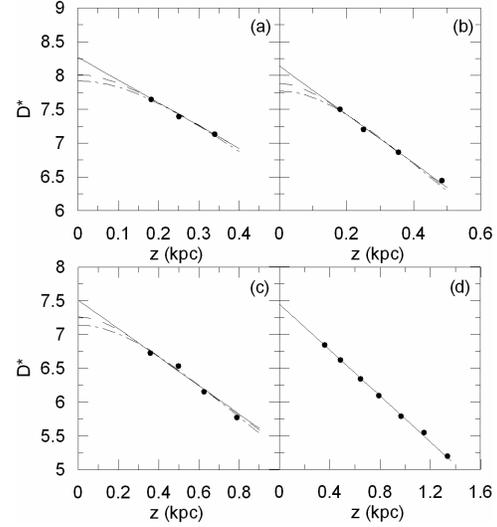}} 
\caption {Comparison of the observed space density function with the density laws 
for different absolute magnitude intervals: (a) (10-11], (b) (9-10], (c) (8-9], 
and (d) (7-8]. The continuous curve represents the exponential law, the dashed 
curve represents the $sech$ law and the dot-dashed curve represents the $sech^{2}$ law 
(the most appropriate density law is indicated in Table 4). The absolute 
magnitudes of these stars are evaluated by the new formula.}
\end {figure}

\begin{figure}
\center
\resizebox{7cm}{10.7cm}{\includegraphics*{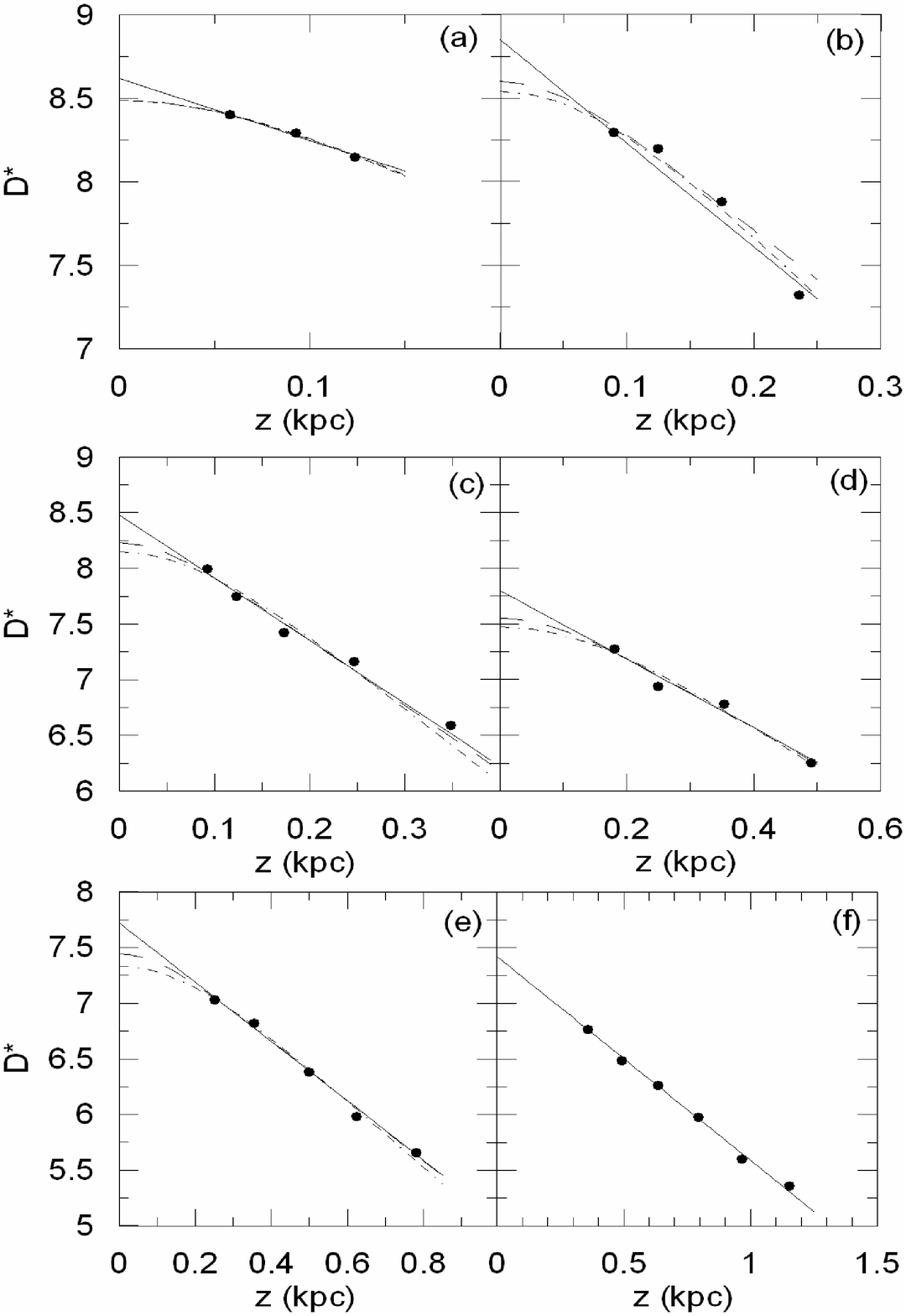}} 
\caption {Comparison of the observed space density function with the density laws 
for different absolute magnitude intervals: (a) (12-13], (b) (11-12], (c) (10-11], 
(d) (9-10], (e) (8-9], and (f) (7-8]. The continuous curve represents the 
exponential law, the dashed curve represents the $sech$ law and the dot-dashed 
curve represents the $sech^{2}$ law (the most appropriate density law is 
indicated in Table 5). The absolute magnitudes of these stars are evaluated by the 
modified formula of Smith et al. (2002).}
\end {figure}

\setcounter{table}{1}
\begin{table*}
\center
\caption{Number of stars as a function of distance $r$ relative to the Sun, 
the corresponding mean distance $z^{*}$ from the Galactic plane, and absolute 
magnitude evaluated by the new formula (equation 5). Distances are in kpc.}
\begin{tabular}{ccccccccccc}
\hline
$M(g) \rightarrow$ &  \multicolumn{2} {c} {(7-8]} & \multicolumn{2} {c} {(8-9]} & 
\multicolumn{2} {c} {(9-10]}& \multicolumn{2} {c} {(10-11]}& \multicolumn{2} {c} {(11-12]}\\
$r_{1}-r_{2}$ & N &  $z^{*}$ & N & $z^{*}$ & N & $z^{*}$ & N & $z^{*}$ & N &  $z^{*}$ \\
\hline
0.10-0.15 &     &       &     &      &    &      &   4 & 0.10 &  1 &  0.10 \\ \cline{8-11}
0.15-0.20 &     &       &     &      &  8 & 0.13 &  12 & 0.13 &  2 &  0.12 \\ \cline{6-7}
0.20-0.30 &   3 &  0.20 &   8 & 0.19 & 49 & 0.18 &  69 & 0.18 &  5 &  0.19 \\ \cline{4-5} \cline{10-11}
0.30-0.40 &  23 &  0.26 &  31 & 0.25 & 48 & 0.25 &  74 & 0.25 & 12 &  0.26 \\ \cline{2-3}\cline{8-9}
0.40-0.60 &  86 &  0.36 &  66 & 0.36 & 90 & 0.36 & 167 & 0.34 & 12 &  0.31 \\
0.60-0.80 & 101 &  0.49 &  82 & 0.50 & 67 & 0.48 &  18 & 0.48 &    &       \\ \cline{6-7}
0.80-1.00 &  87 &  0.64 &  57 & 0.63 & 29 & 0.62 &     &      &    &       \\
1.00-1.25 &  97 &  0.79 &  46 & 0.79 &  5 & 0.78 &     &      &    &       \\ \cline{4-5}
1.25-1.50 &  71 &  0.97 &  22 & 0.97 &    &      &     &      &    &       \\
1.50-1.75 &  57 &  1.05 &   9 & 1.13 &    &      &     &      &    &       \\ 
1.75-2.00 &  34 &  1.34 &     &      &    &      &     &      &    &       \\ \cline{2-3}
2.00-2.50 &  53 &  1.59 &     &      &    &      &     &      &    &       \\
2.50-3.00 &  20 &  1.87 &     &      &    &      &     &      &    &       \\
\hline
    Total & 632 &       & 321 &      & 296 &     & 344 &      & 32 &       \\
\hline
\end{tabular} 
\end{table*}

\setcounter{table}{2}
\begin{table*}
\center
\caption{Number of stars as a function of distance $r$ relative to the Sun, 
the corresponding mean distance $z^{*}$ from the Galactic plane, and absolute 
magnitude evaluated by equation 6. Distances are in kpc.}
\begin{tabular}{rrrrrrrrrrrrrrr}
\hline
$M(g) \rightarrow$  & \multicolumn{2} {c} {(7-8]} & \multicolumn{2} {c} {(8-9]} & 
\multicolumn{2} {c} {(9-10]} &\multicolumn{2} {c} {(10-11]} &\multicolumn{2} {c} {(11-12]} 
& \multicolumn{2} {c} {(12-13]}& \multicolumn{2} {c} {(13-14]} \\
$r_{1}-r_{2}$ & N & $z^{*}$& N & $z^{*}$& N & $z^{*}$& N & $z^{*}$ & N & $z^{*}$ & N & $z^{*}$ & N & $z^{*}$\\
\hline
0.05-0.10 &     &   &   &   &   &  &   &  &     7 &   0.06 &  18 &  0.06 &  1 &  0.06 \\ \cline{10-15}
0.10-0.15 &     &   &   &   &   &  &    19     & 0.09&   38& 0.09 &           38 &       0.09 &         14 &       0.09 \\ \cline{8-9} 
0.15-0.20 &     &   & 2 &  0.13 &  4 &  0.13   & 21  & 0.12& 59 &  0.12 &     53 &       0.12 &         11 &       0.12 \\ \cline{6-7}\cline{12-13}
0.20-0.30 &  1  & 0.20 & 16 & 0.19 & 29 & 0.18 & 41 & 0.17 &117 &  0.18 &     53 &       0.17 &            &            \\ \cline{4-5}\cline{10-11}
0.30-0.40 &  22 & 0.26 & 32 & 0.25 & 26 & 0.25 & 44 & 0.25 & 63 &  0.24 &        &            &            &            \\ \cline{2-3}\cline{8-9}
0.40-0.60 &  72 & 0.36 & 82 & 0.36 & 75 & 0.35 & 48 & 0.35 &  8 &  0.30 &        &            &            &            \\
0.60-0.80 &  74 & 0.49 & 58 & 0.50 & 43 & 0.49 &  5 & 0.47 &    &       &        &            &            &            \\ \cline{6-7}
0.80-1.00 &  73 & 0.64 & 38 & 0.62 & 22 & 0.62 &    &      &    &       &        &            &            &            \\
1.00-1.25 &  73 & 0.79 & 35 & 0.78 & 5  & 0.79 &    &      &    &       &        &            &            &            \\ \cline{4-5}
1.25-1.50 &  46 & 0.97 &  7 & 0.95 &    &      &    &      &    &       &        &            &            &            \\
1.50-1.75 &  37 & 1.15 & 11 & 1.14 &    &      &    &      &    &       &        &            &            &            \\ \cline{2-3}
1.75-2.00 &  35 & 1.33 &  3 & 1.31 &    &      &    &      &    &       &        &            &            &            \\
2.00-2.50 &  34 & 1.54 &    &      &    &      &    &      &    &       &        &            &            &            \\
2.50-3.00 &  12 & 1.99 &    &      &    &      &    &      &    &       &        &            &            &            \\
\hline
Total &479&      &284 &      &204 &      & 178&      &292 &       &    162 &            &         26 &            \\
\hline
\end{tabular}  
\end{table*}

\subsection{Model parameters and luminosity function}

The observed logarithmic space densities are compared with the density laws for 
each absolute magnitude interval, in Fig. 5 and Fig. 6, by $\chi^{2}$-method and 
the resulting Galactic model parameters are given in Table 4 and Table 5, 
respectively. The density law giving the local space density closest to the 
Hipparcos one for a specific absolute magnitude interval has been adopted as the 
most appropriate density law for this comparison and it is indicated in Table 4 
and Table 5 (the details for comparison of the observed space densities with the 
density laws are given in many papers; cf. Karaali et al. 2004b). 
As mentioned in Section 4.1, Table 4 covers stars with absolute magnitudes 
brighter than $M(g)=11$, i.e. $7<M(g)\leq8$, $8<M(g)\leq9$, 
$9<M(g)\leq10$, and $10<M(g)\leq11$. Whereas Table 5 involves two more 
absolute magnitude intervals additional to the ones just cited, i.e. 
$11<M(g)\leq12$ and $12<M(g)\leq13$. We compared the model parameters 
with the ones of thin disk appeared in the literature so far, because Fig. 7 
shows that the spatial distribution of stars for different absolute magnitude 
intervals has a unique modal structure (for detail see our recent paper, Karaali 
et al. 2004b). The range of the scale height for four absolute magnitude 
intervals in Table 4 is 206-354 pc and it is in agreement with the one for thin 
disk appeared in the literature (Karaali et al. 2004b and the references therein; 
see also their Table 1). The same case also holds for the range of scale height 
in Table 5 except two intervals, $11<M(g)\leq12$ and $12<M(g)\leq13$, 
where the scale height is less than 200 pc. The discrepancy between two sets 
of solution is more conspicuous when luminosity function is considered. Actually, 
the luminosity function resulting from the comparison of the observed space 
densities with the density laws in Fig. 5 is rather close to the one of Hipparcos 
(Fig. 8), whereas the luminosity function obtained from the comparison in Fig. 6 
deviates from the luminosity function of Hipparcos at the faint end, i.e. 
$10<M(g)\leq11$, $11<M(g)\leq12$, and $12<M(g)\leq13$, considerably 
(Fig. 9). The bright segment of the luminosity function, i.e. $M(g)<7$ mag, 
is taken from Ak et al. (2005). 

\begin{figure}
\center
\resizebox{7cm}{6.90cm}{\includegraphics*{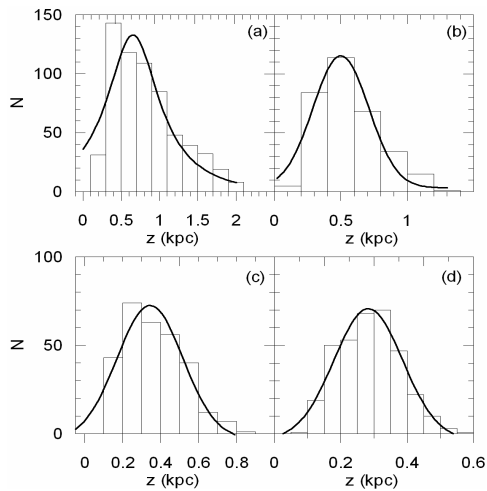}} 
\caption {The spatial distribution for stars whose absolute magnitudes are 
evaluated by the new formula (equation 5) for different absolute magnitude 
intervals: (a) for (7-8], (b) (8-9], (c) (9-10], and (d) (10-11]. The curve 
represents the Gaussian fit.}
\end {figure}

\setcounter{table}{3}
\begin{table}
\center
\caption{Galactic model parameters for different absolute magnitude intervals 
resulting from the comparison of observed logarithmic space densities with a 
(unique) density law (Fig. 5). The columns give: the absolute magnitude 
interval $M(g)$, the density law, the logarithmic local space density 
$n^{*}$, the scale height for $sech^{2}$ density law $z_{0}$, the scale height 
for exponential density law $H$ (pc), $\chi^{2}(10^{-10})$, the standard 
deviation $s$ and the local space densities for $Hipparcos$ $\odot$.}
{\scriptsize
\begin{tabular}{ccccrcc}
\hline
$M(g)$ & Law &   $n^{*}$ & $z_{0}/H$ & $\chi^{2}$ & $s$ & $\odot$\\
\hline
(10-11]& $sech^{2}$ &  7.92$^{+0.02}_{-0.02}$ & 214/354$^{+7}_{-6}$   & 249382 &  0.032 & 7.73\\
 (9-10]& $sech^{2}$ &  7.77$^{+0.04}_{-0.04}$ & 209/346$^{+17}_{-17}$ & 466330 &  0.061 & 7.60\\
  (8-9]& $exp$      &  7.51$^{+0.05}_{-0.05}$ &     206$^{+11}_{-9}$  & 142621 &  0.054 & 7.52\\
  (7-8]& $exp$      &  7.45$^{+0.01}_{-0.01}$ &     256$^{+4}_{-4}$   &  10385 &  0.022 & 7.47\\
\hline
\end{tabular}
}  
\end{table}

\setcounter{table}{4}
\begin{table}
\center
\caption{Galactic model parameters for different absolute magnitude intervals 
resulting from the comparison of observed logarithmic space densities with a 
(unique) density law (Fig. 6). The scale height $z_{0}$ corresponds to either 
$sech$ density law (for $8<M(g)\leq10$) or $sech^{2}$ density law 
(for $10<M(g)\leq13$). The other symbols as in Table 4.}
{\scriptsize
\begin{tabular}{ccccrcc}
\hline
$M(g)$ & Law &    $n^{*}$ &  $z_{0}/H$ & $\chi^{2}$ &        $s$ &   $\odot$ \\
\hline
 (12-13] & $sech^{2}$ & 8.49$^{+0.01}_{-0.01}$ &  89/148$^{+3}_{-4}$   &  146982 &  0.008 & 8.06 \\
 (11-12] & $sech^{2}$ & 8.54$^{+0.06}_{-0.04}$ & 120/199 $^{+20}_{-16}$& 5572217 &  0.070 & 7.91 \\
 (10-11] & $sech^{2}$ & 8.15$^{+0.07}_{-0.07}$ & 130/215 $^{+22}_{-17}$& 5285124 &  0.112 & 7.73 \\
  (9-10] & $sech$     & 7.55$^{+0.07}_{-0.06}$ & 135/224 $^{+21}_{-15}$&  774272 &  0.066 & 7.60 \\
  (8-9]  & $sech$     & 7.44$^{+0.03}_{-0.03}$ & 161/267 $^{+10}_{-8}$ &  118814 &  0.041 & 7.52 \\
  (7-8]  & $exp$      & 7.42$^{+0.01}_{-0.02}$ & 237$^{+4}_{-5}$       &   23702 &  0.035 & 7.47 \\
\hline
\end{tabular}  
}  
\end{table}

The agreement of the luminosity function with the one of Hipparcos is a strong 
evidence for the confidence of the data and the method used in the investigation 
of a field. As the difference between two sets of observed density functions is 
due to only the different procedures in absolute magnitude evaluation, we argue 
that equation (5) supplies significant improvement to the modified equation of 
Smith et al. (2002), i.e. equation (6). 

\begin{figure}
\center
\resizebox{8cm}{5cm}{\includegraphics*{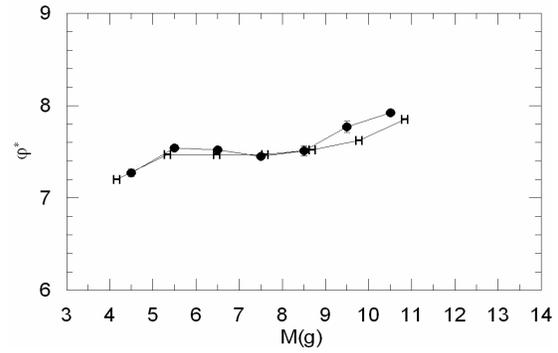}} 
\caption {Luminosity function resulting from the comparison of observed space 
density functions with the density laws in Fig. 5, confronted to the luminosity 
function of \it{Hipparcos} (H).}
\end {figure}

\begin{figure}
\center
\resizebox{8cm}{5cm}{\includegraphics*{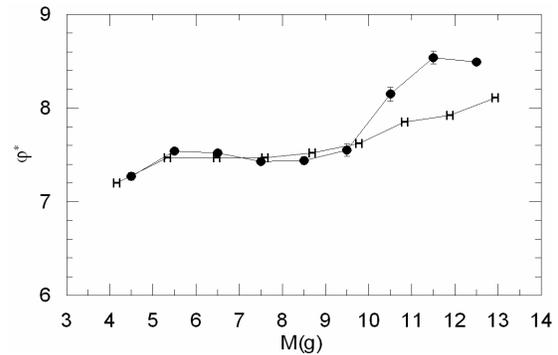}} 
\caption {Luminosity function resulting from the comparison of observed space 
density functions with the density laws in Fig. 6, confronted to the luminosity 
function of \it{Hipparcos} (H).}
\end {figure}

\section{Summary and conclusion}

We presented a new formula for absolute magnitude determination for late-type 
dwarf stars as a function of $(g-r)$ and $(r-i)$ for $Sloan$ 
photometry. This formula has been tested in two ways: (1) the $(g-r)$ 
and $(r-i)$ data transformed from the $(B-V)$ and $(R-I)$ colour indices 
for 708 stars taken from the WEB page of Neill Reid fit with the $(g-r)$ 
and $(r-i)$ data of 6049 stars in a field in the anti-centre direction 
of the Galaxy, and (2) the luminosity function obtained by these absolute 
magnitudes fit with the one of Hipparcos (Jahreiss \& Wielen 1997). Whereas 
the luminosity function corresponding to the absolute magnitudes determined by 
the modified formula of Smith et al. (2002), which extends down to the absolute 
magnitude $M(g)=13$, deviates from the luminosity function of Hipparcos considerably. 
The new approach makes the limiting absolute magnitude brighter, i.e. $M(g)=11$.

The unique-modal spatial distribution of 1625 stars for four absolute magnitude 
intervals shows that this sample belongs to thin disk. Hence, we compared the 
model parameters with the ones for thin disk appeared up to now in the literature 
which have been summarized in the work of Karaali et al. (2004b). The agreement 
is good, so it confirms the data and the method used in their estimation. Also, 
this work confirmed the argue that the model parameters are absolute-magnitude 
dependent, and the density law $sech^{2}$ fits better relative to the exponential 
density law for faint absolute magnitude intervals for thin disk (Table 4).          

\begin{acknowledgements}
We thank the anonymous referees for their suggestions. This research has 
made use of the SIMBAD database, operated at CDS, Strasbourg, France. 
The data was made publically available through the Isaac Newton Groups' 
Wide Field Camera Survey Programme. The Isaac Newton Telescope is operated on 
the island of La Palma by the Isaac Newton Group in the Spanish Observatorio 
del Roque de los Muchachos of the Instituto de Astrofisica de Canarias.  
\end{acknowledgements}

\setcounter{table}{0}
\begin{table*}
\begin{minipage}{185mm} 
\center
\caption{Data for stars used to obtain the new formula for absolute magnitude evaluation. 
The columns give: identification number, equatorial coordinates, magnitude and colours 
for $Sloan$ photometry and the corresponding errors, $UBV$ data and their errors, 
and finally the last column gives the references. The referances are as follows: 
(1) Smith et al. (2002) and (2) Cambridge Astronomical Institute.}
{\tiny
\begin{tabular}{lccccccccccc}
\hline
Star &    $\alpha$    &    $\delta$   &    $g$          &  $(u-g)$       &  $(g-r)$        &  $(r-i)$        & $V$    &  $(U-B)$      &  $(B-V)$       &  $(R-I)$  & Refs\\
\hline
SA 112-805 &   00 42 47 & +00 16 08  & 12.050$\pm$0.003 & 1.220$\pm$0.005 & -0.112$\pm$0.004 &-0.133$\pm$0.004 & 12.086$\pm$0.001&  0.150$\pm$0.003 & 0.152$\pm$0.001 & 0.075$\pm$0.002& 1 \\
SA 92-309  &   00 53 14 & +00 46 02  & 14.080~~~0.007 & 1.170~~~0.018 & 0.370~~~0.009 & 0.115~~~0.010 & 13.842~~~0.004 & -0.024~~~0.003 & 0.513~~~0.006 & 0.325~~~0.004 & 2\\
SA 92-322  &   00 53 47 & +00 47 35  & 12.888~~~0.006 & 1.200~~~0.017 & 0.320~~~0.007 & 0.094~~~0.010 & 12.676~~~0.001 & -0.002~~~0.003 & 0.528~~~0.005 & 0.305~~~0.001 & 2\\
SA 92-248  &   00 54 31 & +00 40 19  & 15.955~~~0.053 & 2.650~~~0.106 & 1.050~~~0.063 & 0.352~~~0.050 & 15.346~~~0.026 &  1.289~~~0.096 & 1.128~~~0.016 & 0.553~~~0.015 & 2\\
SA 92-249  &   00 54 34 & +00 41 06  & 14.637~~~0.012 & 1.520~~~0.025 & 0.500~~~0.014 & 0.162~~~0.010 & 14.325~~~0.005 &  0.240~~~0.011 & 0.699~~~0.008 & 0.370~~~0.007 & 2\\
SA 92-250  &   00 54 37 & +00 38 58  & 13.537~~~0.008 & 1.790~~~0.019 & 0.590~~~0.009 & 0.186~~~0.010 & 13.178~~~0.002 &  0.480~~~0.007 & 0.814~~~0.003 & 0.394~~~0.002 & 2\\
SA 92-330  &   00 54 43 & +00 43 27  & 15.317~~~0.066 & 1.140~~~0.098 & 0.380~~~0.074 & 0.125~~~0.050 & 15.073~~~0.014 & -0.115~~~0.016 & 0.568~~~0.030 & 0.334~~~0.000 & 2\\
SA 92-252  &   00 54 47 & +00 39 25  & 15.171~~~0.012 & 1.090~~~0.023 & 0.370~~~0.014 & 0.122~~~0.010 & 14.932~~~0.003 & -0.140~~~0.008 & 0.517~~~0.005 & 0.332~~~0.007 & 2\\
SA 92-253  &   00 54 51 & +00 40 20  & 14.735~~~0.009 & 2.390~~~0.026 & 1.100~~~0.010 & 0.419~~~0.010 & 14.085~~~0.003 &  0.955~~~0.022 & 1.131~~~0.006 & 0.616~~~0.004 & 2\\
SA 92-335  &   00 54 58 & +00 44 01  & 12.811~~~0.006 & 1.470~~~0.017 & 0.470~~~0.007 & 0.129~~~0.010 & 12.523~~~0.001 &  0.208~~~0.005 & 0.672~~~0.003 & 0.338~~~0.001 & 2\\
SA 92-339  &   00 55 03 & +00 44 12  & 15.803~~~0.027 & 1.000~~~0.043 & 0.330~~~0.031 & 0.129~~~0.030 & 15.579~~~0.009 & -0.177~~~0.013 & 0.449~~~0.012 & 0.339~~~0.020 & 2\\
SA 92-342  &   00 55 10 & +00 43 13  & 11.768~~~0.001 & 1.085~~~0.004 & 0.247~~~0.001 & 0.050~~~0.001 & 11.613~~~0.001 & -0.042~~~0.002 & 0.436~~~0.001 & 0.270~~~0.001 & 1\\
SA 92-188  &   00 55 11 & +00 23 12  & 15.356~~~0.016 & 2.180~~~0.052 & 1.030~~~0.019 & 0.373~~~0.020 & 14.751~~~0.010 &  0.751~~~0.055 & 1.050~~~0.019 & 0.573~~~0.004 & 2\\
SA 92-409  &   00 55 12 & +00 55 58  & 11.292~~~0.008 & 2.540~~~0.020 & 1.130~~~0.009 & 0.428~~~0.010 & 10.627~~~0.003 &  1.136~~~0.009 & 1.138~~~0.003 & 0.625~~~0.003 & 2\\
SA 92-410  &   00 55 14 & +01 01 52  & 15.126~~~0.014 & 1.000~~~0.025 & 0.200~~~0.017 & 0.028~~~0.020 & 14.984~~~0.006 & -0.134~~~0.008 & 0.398~~~ 0.006 & 0.242~~~0.010& 2 \\
SA 92-412  &   00 55 16 & +01 01 56  & 15.233~~~0.017 & 1.030~~~0.030 & 0.290~~~0.020 & 0.093~~~0.020 & 15.036~~~0.005 & -0.152~~~0.013 & 0.457~~~ 0.008 & 0.304~~~0.009& 2 \\
SA 92-259  &   00 55 22 & +00 40 32  & 15.304~~~0.043 & 1.370~~~0.066 & 0.450~~~0.049 & 0.248~~~0.040 & 14.997~~~0.012 & 0.108~~~0.021  & 0.642~~~ 0.022 & 0.452~~~0.020& 2 \\
SA 92-345  &   00 55 24 & +00 51 08  & 15.611~~~0.014 & 1.460~~~0.035 & 0.630~~~0.016 & 0.272~~~0.020 & 15.216~~~0.001 & 0.121~~~0.034  & 0.745~~~ 0.001 & 0.476~~~0.011& 2 \\
SA 92-347  &   00 55 26 & +00 50 50  & 15.999~~~0.070 & 1.140~~~0.106 & 0.390~~~0.082 & 0.108~~~0.100 & 15.752~~~0.026 & -0.097~~~0.036 & 0.543~~~  0.028 & 0.318~~~0.076& 2 \\
SA 92-348  &   00 55 30 & +00 44 34  & 12.367~~~0.007 & 1.300~~~0.017 & 0.400~~~0.008 & 0.131~~~0.010 & 12.109~~~0.001 & 0.056~~~0.004  & 0.598~~~  0.002 & 0.341~~~0.001& 2 \\
SA 92-417  &   00 55 32 & +00 53 08  & 16.176~~~0.039 & 1.020~~~0.063 & 0.410~~~0.047 & 0.094~~~0.080 & 15.922~~~0.013 & -0.185~~~0.032 & 0.477~~~ 0.019 & 0.305~~~0.068& 2 \\
SA 92-260  &   00 55 33 & +00 38 24  & 15.719~~~0.018 & 2.540~~~0.047 & 1.100~~~0.022 & 0.410~~~0.020 & 15.071~~~0.009 & 1.115~~~0.048  & 1.162~~~ 0.009 & 0.608~~~0.006& 2 \\
SA 92-263  &   00 55 39 & +00 36 20  & 12.277~~~0.007 & 2.230~~~0.039 & 0.810~~~0.009 & 0.321~~~0.010 & 11.782~~~0.001 & 0.843~~~0.003  & 1.048~~~ 0.001 & 0.522~~~0.001& 2 \\
SA 92-425  &   00 55 58 & +00 52 59  & 14.625~~~0.008 & 2.610~~~0.021 & 1.170~~~0.010 & 0.429~~~0.010 & 13.941~~~0.004 & 1.173~~~0.012  & 1.191~~~ 0.007 & 0.627~~~0.003& 2 \\
SA 92-426  &   00 56 00 & +00 50 08  & 14.796~~~0.020 & 1.500~~~0.040 & 0.530~~~0.024 & 0.189~~~0.020 & 14.466~~~0.010 & 0.184~~~0.029  & 0.729~~~  0.013 & 0.396~~~0.012& 2 \\
SA 92-355  &   00 56 06 & +00 50 49 & 15.657~~~0.018 & 2.610~~~0.047 & 1.180~~~0.022 & 0.449~~~0.020 & 14.965~~~0.009 & 1.201~~~0.046  & 1.164~~~  0.012 & 0.645~~~0.005& 2 \\
SA 92-430  &   00 56 15 & +00 53 20 & 14.691~~~0.009 & 1.200~~~0.020 & 0.390~~~0.011 & 0.129~~~0.010 & 14.440~~~0.004 &-0.040~~~0.008  & 0.567~~~  0.005 & 0.338~~~0.005& 2 \\
SA 92-276  &   00 56 27 & +00 41 52 & 12.318~~~0.009 & 1.330~~~0.019 & 0.450~~~0.011 & 0.148~~~0.010 & 12.036~~~0.004 & 0.067~~~0.004  & 0.629~~~  0.002 & 0.357~~~0.003& 2 \\
SA 92-282  &   00 56 47 & +00 38 31  & 13.048~~~0.007 & 1.024~~~0.017 & 0.119~~~0.004 & 0.022~~~0.006 & 12.969~~~0.003 & -0.038~~~0.004 & 0.318~~~  0.002 & 0.221~~~0.002& 1 \\
SA 92-507  &   00 56 51 & +01 05 57 & 11.760~~~0.006 & 2.040~~~0.017 & 0.700~~~0.007 & 0.257~~~0.010 & 11.332~~~0.001 & 0.688~~~0.001  & 0.932~~~  0.005 & 0.461~~~0.001& 2 \\
SA 92-508  &   00 56 51 & +01 09 34 & 11.908~~~0.014 & 1.170~~~0.024 & 0.350~~~0.016 & 0.109~~~0.010 & 11.679~~~0.003 & -0.047~~~0.003 & 0.529~~~  0.005 & 0.320~~~0.003& 2  \\
SA 92-364  &   00 56 52 & +00 43 56  & 11.944~~~0.011 & 1.230~~~0.022 & 0.420~~~0.013 & 0.148~~~0.010 & 11.673~~~0.002 & -0.037~~~0.009 & 0.607~~~0.002 & 0.357~~~0.001& 2 \\
SA 92-433  &   00 56 54 & +01 00 43  & 11.946~~~0.008 & 1.380~~~0.021 & 0.440~~~0.009 & 0.139~~~0.010 & 11.667~~~0.001 & 0.110~~~0.012  & 0.655~~~  0.003 & 0.348~~~0.002& 2 \\
SA 92-288  &   00 57 17 & +00 36 49  & 12.019~~~0.006 & 1.760~~~0.013 & 0.668~~~0.005 & 0.233~~~0.003 & 11.630~~~0.001 & 0.472~~~0.002  & 0.855~~~  0.001 & 0.441~~~0.001& 1 \\
SA 93-317  &   01 54 38 & +00 43 01  & 11.741~~~0.003 & 1.081~~~0.008& 0.308~~~0.003  & 0.085~~~0.002 & 11.546~~~0.001 & -0.055~~~0.002 & 0.488~~~  0.001 & 0.298  0.001& 1 \\
SA 93-333  &   01 55 05 & +00 45 43  & 12.399~~~0.003 & 1.736~~~0.006 & 0.639~~~0.004 & 0.203~~~0.003 & 12.011~~~0.002 & 0.436~~~0.003  & 0.832~~~  0.002 & 0.422~~~  0.001& 1 \\
SA 94-242  &   02 57 21 & +00 18 39  & 11.793~~~0.002 & 1.182~~~0.025 & 0.090~~~0.003 &-0.032~~~0.003 & 11.728~~~0.001 & 0.107~~~0.002  & 0.301~~~  0.001 & 0.184~~~0.001& 1 \\
SA 94-702  &   02 58 13 & +01 10 54  & 12.313~~~0.002 & 3.151~~~0.009 & 1.188~~~0.003 & 0.458~~~0.002 & 11.594~~~0.001 & 1.621~~~0.005  & 1.418~~~  0.001 & 0.673~~~0.001& 1 \\
Ross 374   &   03 27 00 & +23 46 36  & 11.022~~~  0.002 & 1.036~~~  0.004 & 0.378~~~  0.003  & 0.141~~~  0.003  & 10.826        & -0.109        & 0.529        &    0.240& 1 \\
SA 95-15   &   03 52 40 &$-$00 05 23  & 11.639~~~  0.006 & 1.460~~~ 0.017 & 0.550~~~  0.007  & 0.177~~~  0.010  & 11.302~~~  0.001 & 0.157~~~  0.004  & 0.712~~~  0.001 & 0.385~~~  0.001& 2 \\
SA 95-16   &   03 52 41 &$-$00 05 06  & 15.044~~~  0.032 & 2.810~~~  0.054 & 1.240~~~  0.037  & 0.480~~~  0.030  & 14.313~~~  0.012 & 1.322~~~  0.031  & 1.306~~~  0.016 & 0.676~~~  0.006& 2 \\
SA 95-301  &   03 52 41 & +00 31 21  & 11.843~~~  0.006 & 2.780~~~  0.017 & 1.050~~~  0.007  & 0.423~~~  0.010  & 11.216~~~  0.001 & 1.296~~~  0.005  & 1.290~~~  0.001 & 0.620~~~  0.001& 2 \\  
SA 95-302  &   03 52 42 & +00 31 17  & 12.081~~~  0.007 & 1.770~~~  0.017 & 0.640~~~  0.008  & 0.214~~~  0.010  & 11.694~~~  0.002 & 0.447~~~  0.006  & 0.825~~~  0.001 & 0.420~~~  0.001& 2  \\
SA 95-96   &   03 52 54 & +00 00 19  & 9.965~~~  0.002 & 1.178~~~  0.004 & -0.094~~~  0.003 & -0.121~~~  0.020 & 10.010~~~  0.002  & 0.072~~~  0.003  & 0.147~~~  0.001 & 0.095~~~  0.001& 1 \\
SA 95-97   &   03 52 58 &$-$00 00 19  & 15.280~~~  0.009 & 1.780~~~  0.030 & 0.730~~~  0.011  & 0.346~~~  0.020  & 14.818~~~  0.001 & 0.380~~~  0.021  & 0.906~~~  0.023 & 0.546~~~  0.019& 2 \\
SA 95-98   &   03 53 00 & +00 02 49  & 15.102~~~  0.020 & 2.540~~~  0.034 & 1.110~~~  0.022  & 0.423~~~  0.020  & 14.448~~~  0.001 & 1.092~~~  0.018  & 1.181~~~  0.001 & 0.620~~~  0.007& 2 \\
SA 95-100  &   03 53 01 & +00 00 15  & 16.078~~~  0.044 & 1.440~~~  0.123 & 0.760~~~  0.056  & 0.215~~~  0.080  & 15.633~~~  0.028 & 0.051~~~  0.113  & 0.791~~~  0.079 & 0.421~~~  0.057& 2 \\
SA 95-101  &   03 53 04 & +00 02 50  & 13.035~~~  0.015 & 1.590~~~  0.027 & 0.570~~~  0.017  & 0.220~~~  0.010  & 12.677~~~  0.003 & 0.263~~~  0.010  & 0.778~~~  0.003 & 0.426~~~  0.006& 2 \\
SA 95-102  &   03 53 08 & +00 01 10  & 16.037~~~  0.044 & 1.690~~~  0.099 & 0.590~~~  0.058  & 0.420~~~  0.080  & 15.622~~~  0.033 & 0.162~~~  0.061  & 1.001~~~  0.080 & 0.618~~~  0.051& 2 \\
SA 95-252  &   03 53 11 & +00 27 24  & 16.160~~~  0.021 & 2.810~~~  0.051 & 1.280~~~  0.024  & 0.555~~~  0.020  & 15.394~~~  0.007 & 1.178~~~  0.043  & 1.452~~~  0.026 & 0.747~~~  0.009& 2 \\
SA 95-190  &   03 53 13 & +00 16 24  & 12.695~~~  0.002 & 1.324~~~  0.004 & 0.109~~~  0.003  & 0.003~~~  0.003  & 12.627~~~ 0.002 & 0.236~~~  0.004  & 0.287~~~  0.002 & 0.220~~~  0.002& 1 \\
SA 95-193  &   03 53 20 & +00 16 35  & 14.975~~~  0.012 & 2.455~~~  0.017 & 1.124~~~  0.006  & 0.406~~~  0.005  & 14.338~~~  0.005 & 1.239~~~  0.026  & 1.211~~~  0.006 & 0.616~~~  0.003& 1 \\
SA 95-105  &   03 53 21 &$-$00 00 18  & 14.058~~~  0.006 & 2.020~~~  0.016 & 0.790~~~  0.007  & 0.335~~~  0.010  & 13.574~~~  0.000 & 0.627~~~  0.000  & 0.976~~~  0.000 &   0.536~~~  0.000& 2 \\
SA 95-106  &   03 53 25 & +00 01 24  & 15.478~~~  0.322 & 2.040~~~  0.458 & 0.490~~~  0.356  & 0.305~~~  0.220  & 15.137~~~  0.006 & 0.369~~~  0.024  & 1.251~~~  0.062 & 0.508~~~  0.013& 2 \\
SA 95-112  &   03 53 40 &$-$00 01 10  & 16.054~~~  0.006 & 1.360~~~  0.016 & 0.890~~~  0.007  & 0.423~~~  0.010  & 15.502~~~  0.000 & 0.077~~~  0.000  & 0.662~~~  0.000 & 0.620~~~  0.000& 2 \\
SA 95-41   &   03 53 41 &$-$00 02 33  & 14.589~~~  0.006 & 1.720~~~  0.016 & 0.860~~~  0.007  & 0.386~~~  0.010  & 14.060~~~  0.000 & 0.297~~~  0.000  & 0.903~~~  0.000 &   0.585~~~  0.000& 2 \\
SA 95-317  &   03 53 44 & +00 29 51  & 14.164~~~  0.010 & 2.660~~~  0.022 & 1.190~~~  0.012  & 0.514~~~  0.010  & 13.449~~~  0.003 & 1.120~~~  0.013  & 1.320~~~  0.007 & 0.708~~~  0.001& 2 \\
SA 95-263  &   03 53 47 & +00 26 38  & 13.423~~~  0.008 & 3.140~~~  0.020 & 1.250~~~  0.009  & 0.517~~~  0.010  & 12.679~~~  0.003 & 1.559~~~  0.009  & 1.500~~~  0.003 & 0.711~~~  0.001& 2 \\
SA 95-115  &   03 53 48 &$-$00 00 48  & 15.198~~~  0.006 & 1.510~~~  0.016 & 0.840~~~  0.007  & 0.380~~~  0.010  & 14.680~~~  0.000 & 0.096~~~  0.000  & 0.836~~~  0.000 &   0.579~~~  0.000& 2 \\
SA 95-43   &   03 53 49 &$-$00 03 00  & 11.023~~~  0.008 & 1.170~~~  0.019 & 0.330~~~  0.009  & 0.105~~~  0.010  & 10.803~~~  0.002 & -0.016~~~  0.003 & 0.510~~~  0.002 & 0.316~~~ 0.002& 2 \\
SA 95-271  &   03 54 16 & +00 18 53  & 14.357~~~  0.009 & 2.480~~~  0.022 & 1.130~~~  0.011  & 0.523~~~  0.010  & 13.669~~~  0.006 & 0.916~~~  0.013  & 1.287~~~  0.008 & 0.717~~~  0.002& 2 \\
SA 95-329  &   03 54 24 & +00 37 08  & 15.314~~~  0.012 & 2.540~~~  0.033 & 1.190~~~  0.014  & 0.445~~~  0.020  & 14.617~~~  0.005 & 1.093~~~  0.031  & 1.184~~~  0.010 & 0.642~~~  0.009& 2 \\
SA 95-276  &   03 54 46 & +00 25 55  & 14.800~~~  0.012 & 2.670~~~  0.029 & 1.150~~~  0.014  & 0.449~~~  0.010  & 14.118~~~  0.006 & 1.218 ~~~ 0.022  & 1.225~~~  0.010 & 0.646~~~  0.003& 2 \\
SA 95-60   &   03 54 50 &$-$00 07 02  & 13.817~~~  0.009 & 1.540~~~  0.019 & 0.620~~~  0.010  & 0.244~~~  0.010  & 13.429~~~  0.003 & 0.197~~~  0.006  & 0.776~~~  0.003 & 0.449~~~  0.002& 2 \\
SA 95-218  &   03 54 50 & +00 10 10  & 12.399~~~  0.002 & 1.444~~~  0.007 & 0.500~~~  0.000  & 0.167~~~  0.003  & 12.095~~~ 0.003 & 0.208~~~  0.003  & 0.708~~~  0.002 & 0.370~~~  0.002& 1 \\
SA 95-132  &   03 54 52 & +00 05 26  & 12.229~~~  0.004 & 1.498~~~  0.007 & 0.240~~~  0.005  & 0.079~~~  0.004  & 12.064~~~  0.002 & 0.300~~~  0.006  & 0.448~~~  0.002 & 0.287~~~  0.002& 1 \\
SA 95-62   &   03 55 00 &$-$00 02 54  & 14.225~~~  0.008 & 2.740 ~~~ 0.021 & 1.140~~~  0.009  & 0.490~~~ 0.010   & 13.538~~~  0.003 & 1.181~~~  0.014  & 1.355~~~  0.005 & 0.685~~~  0.002& 2 \\
SA 95-139  &   03 55 05 & +00 03 11  & 12.675~~~  0.008 & 2.020~~~  0.023 & 0.810~~~  0.009  & 0.272~~~  0.010  & 12.196~~~  0.002 & 0.677~~~  0.019  & 0.923~~~  0.005 & 0.476~~~  0.002& 2 \\
SA 95-66   &   03 55 06 &$-$00 09 30  & 13.244~~~  0.006 & 1.470~~~  0.018 & 0.550~~~  0.007  & 0.232~~~  0.010  & 12.892~~~  0.002 & 0.167~~~  0.004  & 0.715~~~  0.007 & 0.438~~~  0.006& 2 \\
SA 95-142  &   03 55 09 & +00 01 22  & 13.191~~~  0.008 & 1.278~~~  0.011 & 0.418~~~  0.006  & 0.163~~~  0.007  & 12.927~~~  0.003 & 0.097~~~  0.004  & 0.588~~~  0.003 & 0.375~~~  0.002& 1 \\
SA 95-227  &   03 55 09 & +00 14 35  & 16.236~~~  0.028 & 1.410~~~ 0.056 & 0.720~~~  0.033  & 0.351~~~  0.030  & 15.779~~~  0.012 & 0.034~~~  0.042  & 0.771~~~  0.029 & 0.552~~~  0.011& 2 \\
SA 95-74   &   03 55 31 &$-$00 09 14  & 12.066~~~  0.006 & 2.180~~~  0.017 & 0.880~~~  0.007  & 0.367~~~  0.010  & 11.531~~~  0.002 & 0.686~~~  0.004  & 1.126~~~  0.001 & 0.567~~~  0.001& 2 \\
SA 95-231  &   03 55 39 & +00 10 43  & 14.396~~~  0.012 & 1.370~~~  0.023 & 0.260~~~  0.014  & 0.078~~~  0.010  & 14.216~~~  0.004 & 0.297~~~  0.007  & 0.452~~~  0.005 & 0.290~~~  0.005& 2 \\
SA 95-284  &   03 55 42 & +00 26 39  & 14.441~~~  0.009 & 2.690~~~  0.027 & 1.290~~~  0.011  & 0.574~~~  0.010  & 13.669~~~  0.004 & 1.073 ~~~ 0.024  & 1.398~~~  0.008 & 0.766~~~  0.004& 2 \\
SA 95-285  &   03 55 44 & +00 25 11  & 16.110~~~  0.018 & 2.050~~~  0.060 & 0.890~~~  0.021  & 0.403~~~ 0.020  & 15.561~~~  0.007 & 0.703~~~  0.064  & 0.937~~~  0.025 & 0.602~~~  0.006& 2 \\
SA 95-236  &   03 56 13 & +00 08 48  & 11.831~~~  0.006 & 1.480~~~  0.017 & 0.540~~~  0.007  & 0.205~~~  0.010  & 11.491~~~  0.001 & 0.162~~~  0.004  & 0.736~~~  0.001 & 0.411~~~  0.001& 2 \\
BD-21 0910 &   04 33 16 &$-$21 08 07  & 10.079~~~  0.002 & 1.367~~~  0.006 & 0.473~~~  0.003  & 0.143~~~0.003  &  9.796~~~0.008      & 0.139~~~  0.018  & 0.680~~~  0.011 &     0.300& 1\\
SA 96-36   &   04 51 42 &$-$00 10 09  & 10.611~~~  0.002 & 1.226~~~  0.004 & 0.007~~~  0.003  & -0.077~~~  0.003 & 10.591~~~0.001  & 0.118~~~  0.003  & 0.247~~~  0.001 & 0.136~~~  0.001& 1 \\
SA 96-737  &   04 52 35 & +00 22 29  & 12.408~~~  0.003 & 2.736~~~  0.009 & 1.123~~~  0.004  & 0.490~~~  0.004  & 11.716~~~0.002    & 1.160~~~  0.005  & 1.334~~~  0.002 & 0.695~~~  0.001& 1 \\
SA 96-83   &   04 52 59 &$-$00 14 44  & 11.704~~~  0.003 & 1.272~~~  0.007 & -0.078~~~  0.004 & -0.108~~~  0.004  & 11.719~~~0.001  & 0.202~~~  0.005  & 0.179~~~  0.001 & 0.097~~~  0.001& 1 \\
\end{tabular}  
}
\end{minipage}
\end{table*}

\begin{table*}
\begin{minipage}{185mm} 
{\tiny
\begin{tabular}{lccccccccccc}
\hline
Star &    $\alpha$    &    $\delta$   &    $g$          &  $(u-g)$       &  $(g-r)$        &  $(r-i)$        & $V$    &  $(U-B)$      &  $(B-V)$       &  $(R-I)$& Refs. \\
\hline
Ross 49    &   05 44 57 & +09 14 31  & 11.626$\pm$0.002 & 1.134$\pm$0.004 & 0.465$\pm$0.002  & 0.162$\pm$0.001  & 11.817        & -0.180        & 0.560        &   0.263 & 1    \\
SA 97-249  &   05 57 08 & +00 01 12  & 12.013~~~  0.002 & 1.322~~~  0.005 & 0.448~~~  0.002  & 0.141~~~  0.001  & 11.733~~~0.001& 0.100~~~  0.001  & 0.648~~~  0.001 & 0.353~~~  0.001& 1 \\
SA 97-351  &   05 57 37 & +00 13 44  &  9.790~~~  0.003 & 1.161~~~  0.004 & -0.013~~~  0.004  & -0.072~~~  0.003 &  9.781~~~0.001& 0.096~~~  0.002  & 0.202~~~  0.001 & 0.141~~~  0.001& 1 \\
SA 97-75   &   05 57 55 &$-$00 09 29  & 12.466~~~  0.003 & 3.813~~~  0.017 & 1.664~~~  0.004  & 0.753~~~  0.003 & 11.483~~~0.004& 2.100~~~  0.010  & 1.872~~~  0.005 & 0.952~~~  0.002& 1 \\
SA 97-288  &   05 58 30 & +00 06 41  & 11.111~~~  0.001 & 1.339~~~  0.003 & 0.397~~~  0.002  & 0.107~~~  0.003  & 10.870        & 0.130         & 0.600        &    0.270& 1    \\
Hilt 566   &   06 32 10 & +03 34 44  & 11.469~~~  0.002 & 1.116~~~  0.004 & 0.677~~~  0.003  & 0.340~~~  0.003  & 11.140        & -0.100        & 0.820        &    0.330& 1    \\
SA 98-562  &   06 51 31 &$-$00 19 00  & 12.399~~~  0.007 & 1.190~~~  0.017 & 0.330~~~  0.009  & 0.092~~~  0.010  & 12.185~~~  0.003 & -0.002~~~  0.004 & 0.522~~~  0.005 & 0.303~~~  0.001& 2 \\
SA 98-563  &   06 51 32 &$-$00 26 26  & 14.370~~~  0.012 & 0.970~~~  0.023 & 0.310~~~  0.014  & 0.107~~~  0.010  & 14.162~~~  0.005 & -0.190~~~  0.007 & 0.416~~~  0.009 & 0.317~~~  0.008& 2 \\
SA 98-978  &   06 51 34 &$-$00 11 32  & 10.814~~~  0.014 & 1.284~~~  0.005 & 0.404~~~  0.002  & 0.107~~~  0.002  & 10.572~~~  0.001 & 0.094~~~  0.002  & 0.609~~~  0.001 & 0.322~~~  0.001& 1 \\
SA 98-L1   &   06 51 39 &$-$00 26 35  & 16.355~~~  0.026 & 2.340~~~  0.091 & 1.120~~~  0.030  & 0.518~~~  0.030  & 15.672~~~  0.008 & 0.776~~~  0.098  & 1.243~~~  0.046 & 0.712~~~  0.025& 2 \\
SA 98-580  &   06 51 40 &$-$00 27 00  & 14.887~~~  0.039 & 1.310~~~  0.060 & 0.210~~~  0.050  & 0.094~~~  0.060  & 14.728~~~  0.026 & 0.303~~~  0.019  & 0.367~~~  0.019 & 0.305~~~  0.046& 2 \\
SA 98-581  &   06 51 40 &$-$00 26 00  & 14.593~~~  0.048 & 1.100~~~  0.075 & -0.020~~~  0.058 & 0.031~~~  0.050  & 14.556~~~  0.025 & 0.161~~~  0.020  & 0.238~~~  0.028 & 0.244~~~  0.020& 2 \\
SA 98-L4   &   06 51 42 &$-$00 16 20  & 17.212~~~  0.073 & 2.660~~~  0.284 & 1.510~~~  0.096  & 0.594~~~  0.100  & 16.332~~~  0.058 & 1.086~~~  0.343  & 1.344~~~  0.033 & 0.785~~~  0.037& 2 \\
SA 98-1002 &   06 51 43 &$-$00 16 00  & 14.843~~~  0.017 & 1.220~~~  0.029 & 0.420~~~  0.020  & 0.171~~~  0.020  & 14.568~~~  0.006 & -0.027~~~  0.011 & 0.574~~~  0.007 & 0.379~~~  0.013& 2 \\
SA 98-590  &   06 51 43 &$-$00 22 00  & 15.353~~~  0.016 & 2.490~~~  0.049 & 1.160~~~  0.021  & 0.555~~~  0.020  & 14.642~~~  0.011 & 0.853~~~  0.053  & 1.352~~~  0.012 & 0.747~~~  0.011& 2 \\
SA 98-614  &   06 51 49 &$-$00 20 33  & 16.431~~~  0.065 & 1.920~~~  0.102 & 1.320~~~  0.081  & 0.449~~~  0.080  & 15.674~~~  0.042 & 0.399~~~  0.031  & 1.063~~~  0.047 & 0.645~~~  0.037& 2 \\
SA 98-624  &   06 51 52 &$-$00 20 16  & 14.147~~~  0.016 & 1.700~~~  0.032 & 0.540~~~  0.022  & 0.197~~~  0.030  & 13.811~~~  0.014 & 0.394~~~  0.004  & 0.791~~~  0.024 & 0.404~~~  0.021& 2 \\
SA 98-627  &   06 51 53 &$-$00 22 02  & 15.241~~~  0.009 & 1.380~~~  0.024 & 0.560~~~  0.012  & 0.180~~~  0.020  & 14.900~~~  0.006 & 0.078~~~  0.008  & 0.689~~~  0.017 & 0.387~~~  0.013& 2 \\
SA 98-634  &   06 51 56 &$-$00 20 53  & 14.906~~~  0.025 & 1.390~~~  0.040 & 0.470~~~  0.028  & 0.164~~~  0.030  & 14.608~~~  0.004 & 0.123~~~  0.013  & 0.647~~~  0.005 & 0.372~~~  0.018& 2 \\
SA 98-642  &   06 51 59 &$-$00 21 33  & 15.523~~~  0.046 & 1.480~~~  0.076 & 0.320~~~  0.054  & 0.186~~~  0.040  & 15.290~~~  0.019 & 0.318~~~  0.012  & 0.571~~~  0.045 & 0.393~~~  0.002& 2 \\
SA 98-185  &   06 52 02 &$-$00 27 22  & 10.539~~~  0.006 & 1.187~~~  0.008 & -0.060~~~  0.005 & -0.091~~~  0.004 & 10.536~~~  0.002 & 0.113~~~  0.003  & 0.202~~~  0.001 & 0.124~~~  0.001& 1 \\
SA 98-193  &   06 52 03 &$-$00 27 19  & 10.585~~~  0.008 & 2.770~~~  0.103 & 0.910~~~  0.011  & 0.341~~~  0.010  & 10.030~~~  0.002 & 1.152~~~  0.002  & 1.180~~~  0.001 & 0.537~~~  0.001& 2 \\
SA 98-653  &   06 52 05 &$-$00 18 18  &  9.424~~~  0.006 & 0.720~~~  0.016 & -0.230~~~  0.007 & -0.216~~~  0.010 &  9.539~~~  0.001 & -0.099~~~  0.001 & -0.004~~~  0.000 & 0.008~~~  0.001& 2\\
SA 98-650  &   06 52 05 &$-$00 19 39  & 12.237~~~  0.007 & 1.000~~~  0.017 & -0.090~~~  0.008 & -0.134~~~  0.010 & 12.271~~~  0.002 & 0.110~~~  0.004  & 0.157~~~  0.001 & 0.086~~~  0.002& 2 \\
SA 98-652  &   06 52 05 &$-$00 21 57  & 15.015~~~  0.097 & 1.360~~~  0.140 & 0.270~~~  0.108  & 0.129~~~  0.070  & 14.817~~~  0.011 & 0.126~~~  0.018  & 0.611~~~  0.030 & 0.339~~~  0.024& 2 \\
SA 98-666  &   06 52 10 &$-$00 24 00  & 12.712~~~  0.011 & 0.920~~~  0.021 & -0.070~~~  0.013 & -0.111~~~  0.010 & 12.732~~~  0.003 & -0.004~~~  0.004 & 0.164~~~  0.003 & 0.108~~~  0.003& 2 \\
SA 98-670  &   06 52 12 &$-$00 19 00  & 12.592~~~  0.007 & 2.840~~~  0.017  & 1.110~~~  0.008 & 0.457~~~  0.010  & 11.930~~~  0.002 & 1.313~~~  0.006  & 1.356~~~  0.002 & 0.653~~~  0.001& 2 \\
SA 98-671  &   06 52 12 &$-$00 18 30  & 13.880~~~  0.010 & 2.090~~~  0.021  & 0.830~~~  0.012 & 0.291~~~  0.010  & 13.385~~~  0.004 & 0.719~~~  0.011  & 0.968~~~  0.005 & 0.494~~~  0.003& 2 \\
SA 98-676  &   06 52 14 &$-$00 19 00  & 13.701~~~  0.009 & 2.180~~~  0.021  & 1.030~~~  0.011 & 0.478~~~  0.030  & 13.068~~~  0.003 & 0.666~~~  0.011  & 1.146~~~  0.004 & 0.673~~~  0.022& 2 \\
SA 98-682  &   06 52 17 &$-$00 20 00  & 14.028~~~  0.008 & 1.360~~~  0.019  & 0.440~~~  0.009 & 0.143~~~  0.010  & 13.749~~~  0.004 & 0.098~~~  0.006  & 0.632~~~  0.004 & 0.352~~~  0.003& 2 \\
SA 98-685  &   06 52 18 &$-$00 20 20  & 12.129~~~  0.006 & 1.233~~~  0.010  & 0.277~~~  0.004 & 0.071~~~  0.004  & 11.954~~~  0.003 & 0.096~~~  0.003  & 0.463~~~  0.002 & 0.280~~~  0.002& 1 \\
SA 98-688  &   06 52 19 &$-$00 24 00  & 12.810~~~  0.010 & 1.210~~~  0.021  & 0.050~~~  0.012 & -0.036~~~  0.010 & 12.754~~~  0.003 & 0.245~~~  0.008  & 0.293 ~~~ 0.002 & 0.180~~~  0.005& 2 \\
SA 98-1082 &   06 52 20 &$-$00 14 15  & 15.457~~~  0.011 & 1.440~~~  0.029  & 0.660~~~  0.014 & 0.421~~~  0.020  & 15.010~~~  0.006 & -0.001~~~  0.023 & 0.835~~~  0.014 & 0.619~~~  0.013& 2 \\
SA 98-1102 &   06 52 29 &$-$00 14 00  & 12.203~~~  0.009 & 1.110~~~  0.019  & 0.120~~~  0.010 & -0.020~~~  0.010 & 12.113~~~  0.003 & 0.089~~~  0.006  & 0.314~~~  0.003 & 0.195~~~  0.004& 2 \\
SA 98-1112 &   06 52 35 &$-$00 15 00  & 14.340~~~  0.014 & 1.640~~~  0.027  & 0.580~~~  0.017 & 0.225~~~  0.010  & 13.975~~~  0.007 & 0.286~~~  0.015  & 0.814~~~  0.004 & 0.431~~~  0.003& 2 \\
SA 98-1119 &   06 52 37 &$-$00 14 00  & 12.097~~~  0.007 & 1.270~~~  0.017  & 0.340~~~  0.009 & 0.088~~~  0.010  & 11.878~~~  0.002 & 0.069~~~  0.004  & 0.551~~~  0.004 & 0.299~~~  0.004& 2 \\
SA 98-724  &   06 52 37 &$-$00 19 21  & 11.621~~~  0.008 & 2.330~~~  0.019  & 0.830~~~  0.009 & 0.325~~~  0.010  & 11.118~~~  0.003 & 0.904~~~  0.005  & 1.104~~~  0.003 & 0.527~~~  0.002& 2 \\
SA 98-1122 &   06 52 37 &$-$00 17 05  & 14.400~~~  0.010 & 1.020~~~  0.021  & 0.460~~~  0.012 & 0.237~~~  0.010  & 14.090~~~  0.003 & -0.297~~~  0.007 & 0.595~~~  0.006 & 0.442~~~  0.003& 2 \\
SA 98-1124 &   06 52 38 &$-$00 17 00  & 13.781~~~  0.009 & 1.240~~~  0.020  & 0.080~~~  0.011 & -0.015~~~  0.010 & 13.707~~~  0.003 & 0.258~~~  0.008  & 0.315~~~  0.004 & 0.201~~~  0.005& 2 \\
SA 98-733  &   06 52 40 &$-$00 17 16  & 12.878~~~  0.009 & 2.610~~~  0.020  & 1.060~~~  0.011 & 0.454~~~  0.010  & 12.238~~~  0.003 & 1.087~~~  0.006  & 1.285~~~  0.004 & 0.650~~~  0.002& 2 \\
Ru 149F    &   07 24 14 &$-$00 31 39  & 14.006~~~  0.002 & 2.449~~~  0.013  & 0.883~~~  0.003 & 0.316~~~  0.002  & 13.471~~~0.003   & 1.025~~~  0.022  & 1.115~~~  0.008 & 0.538~~~  0.002& 1 \\
Ru 149B    &   07 24 17 &$-$00 33 05  & 12.930~~~  0.002 & 1.389~~~  0.006  & 0.458~~~  0.002 & 0.141~~~  0.001  & 12.642~~~0.002   & 0.151~~~  0.005  & 0.662~~~  0.003 & 0.354~~~  0.003& 1 \\
SA 100-241 &   08 52 34 &$-$00 39 46  & 10.100~~~  0.003 & 1.201~~~  0.005  & -0.092~~~ 0.004 & -0.126~~~  0.004 & 10.139~~~0.001   & 0.101~~~  0.002  & 0.157~~~  0.001 & 0.085~~~  0.001& 1 \\
SA 100-280 &   08 53 35 &$-$00 36 39  & 11.984~~~  0.002 & 1.156~~~  0.006  & 0.299~~~  0.003 & 0.085~~~  0.003  & 11.799~~~0.001   & -0.002~~~  0.001 & 0.494~~~  0.001 & 0.291~~~  0.001& 1 \\
BD -12 2918 &   09 31 18 &$-$13 29 20  & 10.850~~~  0.002 & 2.769~~~  0.005  & 1.339~~~  0.003 & 1.195~~~  0.004  & 10.074~~~0.010   & 1.153~~~  0.013  & 1.535~~~  0.014 & 1.337~~~  0.010& 1 \\
SA 101-315 &   09 54 51 &$-$00 27 28  & 11.798~~~  0.007 & 2.500~~~  0.030  & 0.900~~~  0.014 & 0.361~~~  0.020  & 11.249~~~  0.004 & 1.056~~~  0.007  & 1.153~~~  0.002 & 0.559~~~  0.001& 2 \\
SA 101-316 &   09 54 52 &$-$00 18 34  & 11.779~~~  0.007 & 1.165~~~  0.005  & 0.301~~~  0.001 & 0.074~~~  0.001  & 11.552~~~  0.003 & 0.032~~~  0.004  & 0.493~~~  0.002 & 0.291~~~  0.003& 1 \\
SA 101-L1  &   09 55 29 &$-$00 21 43  & 16.870~~~  0.061 & 1.300~~~  0.094  & 0.540~~~  0.075 & 0.325~~~  0.110  & 16.501~~~ 0.031 & -0.104~~~  0.029 & 0.757~~~  0.034 & 0.527~~~  0.086 & 2\\
SA 101-320 &   09 55 33 &$-$00 22 32  & 14.340~~~  0.011 & 2.130~~~  0.026  & 0.840~~~  0.013 & 0.361~~~  0.010  & 13.823~~~  0.005 & 0.690~~~  0.018  & 1.052~~~  0.010 & 0.561~~~  0.004& 2 \\
SA 101-404 &   09 55 41 &$-$00 18 20  & 13.916~~~  0.013 & 2.090~~~  0.025  & 0.750~~~  0.015 & 0.297~~~  0.010  & 13.459~~~  0.004 & 0.697~~~  0.010  & 0.996~~~  0.008 & 0.500~~~  0.003& 2 \\
SA 101-324 &   09 55 57 &$-$00 23 13  & 10.257~~~  0.006 & 2.560~~~  0.017  & 0.860~~~  0.007 & 0.317~~~  0.010  & 9.742~~~  0.001  & 1.148~~~  0.002  & 1.161~~~  0.001 & 0.519~~~  0.001& 2 \\
SA 101-408 &   09 56 08 &$-$00 12 41  & 15.431~~~  0.044 & 2.750~~~  0.073  & 1.100~~~  0.053 & 0.404~~~  0.040  & 14.785~~~  0.024 & 1.347~~~  0.017  & 1.200~~~  0.044 & 0.603~~~  0.012& 2 \\
SA 101-326 &   09 56 08 &$-$00 27 11  & 15.242~~~  0.015 & 1.530~~~  0.028  & 0.520~~~  0.018 & 0.167~~~  0.020  & 14.923~~~  0.008 & 0.227~~~  0.013  & 0.729~~~  0.009 & 0.375~~~  0.011& 2 \\
SA 101-262 &   09 56 08 &$-$00 29 49  & 14.647~~~  0.018 & 1.620~~~  0.035  & 0.580~~~  0.021 & 0.180~~~  0.020  & 14.295~~~  0.005 & 0.297~~~  0.021  & 0.784~~~  0.016 & 0.387~~~  0.010& 2 \\
SA 101-410 &   09 56 09 &$-$00 14 02  & 13.859~~~  0.025 & 1.170~~~  0.038  & 0.310~~~  0.029 & 0.116~~~  0.020  & 13.646~~~  0.008 & -0.063~~~  0.002 & 0.546~~~  0.001 & 0.326~~~  0.004& 2 \\
SA 101-327 &   09 56 09 &$-$00 25 51  & 14.079~~~  0.008 & 2.550~~~  0.020  & 1.100~~~  0.009 & 0.374~~~  0.010  & 13.441~~~  0.003 & 1.139~~~  0.010  & 1.155~~~  0.006 & 0.574~~~  0.002& 2 \\
SA 101-413 &   09 56 14 &$-$00 11 55  & 13.039~~~  0.009 & 2.100~~~  0.021  & 0.740~~~  0.010 & 0.295~~~  0.010  & 12.583~~~  0.002 & 0.716~~~  0.010  & 0.983~~~  0.005 & 0.497~~~  0.004& 2 \\
SA 101-330 &   09 56 21 &$-$00 27 21  & 13.981~~~  0.009 & 1.220~~~  0.019  & 0.400~~~  0.011 & 0.129~~~  0.010  & 13.723~~~  0.004 & -0.026~~~  0.005 & 0.577~~~ 0.004 & 0.338~~~  0.006& 2 \\
SA 101-415 &   09 56 23 &$-$00 16 54  & 15.520~~~  0.043 & 1.230~~~  0.065  & 0.400~~~  0.048 & 0.141~~~  0.030  & 15.259~~~  0.003 & -0.008~~~  0.002 & 0.577~~~  0.025 & 0.350~~~  0.013& 2 \\
SA 101-270 &   09 56 27 &$-$00 35 41  & 13.949~~~  0.012 & 1.260~~~  0.024  & 0.380~~~  0.013 & 0.095~~~  0.010  & 13.711~~~  0.003 & 0.055~~~  0.013  & 0.554~~~  0.001 & 0.306~~~  0.007& 2 \\
SA 101-278 &   09 56 54 &$-$00 29 37  & 16.020~~~  0.046 & 2.160~~~  0.072  & 0.870~~~  0.053 & 0.347~~~  0.040  & 15.494~~~  0.018 & 0.737~~~  0.027  & 1.041~~~  0.023 & 0.548~~~  0.015& 2 \\
SA 101-L3  &   09 56 55 &$-$00 30 24  & 16.268~~~  0.085 & 1.260~~~  0.134  & 0.500~~~  0.097 & 0.188~~~  0.080  & 15.953~~~  0.023 & -0.033~~~  0.071 & 0.637~~~  0.018 & 0.395~~~  0.049& 2 \\
SA 101-281 &   09 57 05 &$-$00 31 42  & 11.943~~~  0.008 & 1.740~~~  0.019  & 0.600~~~  0.009 & 0.206~~~  0.010  & 11.575~~~  0.004 & 0.419~~~  0.005  & 0.812~~~  0.002 & 0.412~~~  0.001& 2 \\
SA 101-L4  &   09 57 08 &$-$00 31 25  & 16.656~~~  0.078 & 1.680~~~  0.112  & 0.840~~~  0.104 & -0.160~~~  0.290 & 16.264~~~  0.017 & 0.362~~~  0.001  & 0.793~~~  0.014 & 0.062~~~  0.264& 2 \\
SA 101-L5  &   09 57 11 &$-$00 30 40  & 16.237~~~  0.021 & 1.360~~~  0.077  & 0.530~~~  0.025 & 0.094~~~  0.020  & 15.928~~~  0.011 & 0.115~~~  0.086  & 0.622~~~  0.028 & 0.305~~~  0.013& 2 \\
SA 101-421 &   09 57 16 &$-$00 17 17  & 13.411~~~  0.008 & 1.160~~~  0.018  & 0.370~~~  0.009 & 0.084~~~  0.010  & 13.180~~~  0.002 & -0.031~~~  0.004 & 0.507~~~  0.002 & 0.296~~~  0.001& 2 \\
SA 101-338 &   09 57 18 &$-$00 21 02  & 14.050~~~  0.010 & 1.300~~~  0.021  & 0.410~~~  0.013 & 0.130~~~  0.010  & 13.788~~~  0.007 & 0.024~~~  0.004  & 0.634~~~  0.007 & 0.340~~~ 0.004& 2  \\
SA 101-339 &   09 57 18 &$-$00 25 03  & 14.819~~~  0.052 & 1.830~~~  0.076  & 0.610~~~  0.058 & 0.191~~~  0.040  & 14.449~~~  0.010 & 0.501~~~ 0.010   & 0.850~~~  0.008 & 0.398~~~  0.004& 2 \\
SA 101-424 &   09 57 20 &$-$00 16 26  & 15.409~~~  0.056 & 1.590~~~  0.087  & 0.560~~~  0.067 & 0.219~~~  0.060  & 15.058~~~  0.028 & 0.273~~~  0.019  & 0.764~~~  0.039 & 0.425~~~  0.021& 2 \\
SA 101-427 &   09 57 26 &$-$00 17 17  & 15.349~~~  0.059 & 1.660~~~  0.088  & 0.660~~~  0.070 & 0.160~~~  0.050  & 14.964~~~  0.027 & 0.321~~~  0.001  & 0.805~~~  0.030 & 0.369~~~  0.011& 2 \\
SA 101-342 &   09 57 31 &$-$00 21 50  & 15.828~~~  0.053 & 1.150~~~  0.077  & 0.390~~~  0.060 & 0.213~~~  0.040  & 15.556~~~  0.014 & -0.065~~~  0.008 & 0.529~~~  0.008 & 0.419~~~  0.002& 2 \\
SA 101-341 &   09 57 31 &$-$00 21 54  & 14.581~~~  0.013 & 1.280~~~  0.026  & 0.380~~~  0.018 & 0.099~~~  0.020  & 14.342~~~  0.011 & 0.059~~~  0.007  & 0.575~~~  0.014 & 0.309~~~  0.008& 2 \\
SA 101-343 &   09 57 31 &$-$00 22 56  & 15.806~~~  0.056 & 1.330~~~  0.090  & 0.500~~~  0.067 & 0.129~~~  0.060  & 15.504~~~  0.028 & 0.094~~~  0.018  & 0.606~~~  0.049 & 0.338~~~  0.025& 2 \\
SA 101-429 &   09 57 32 &$-$00 18 14  & 14.035~~~  0.008 & 2.150~~~  0.019  & 0.910~~~  0.009 & 0.325~~~  0.010  & 13.496~~~  0.004 & 0.782~~~  0.006  & 0.980~~~  0.002 & 0.526~~~  0.002& 2 \\
SA 101-431 &   09 57 37 &$-$00 17 54  & 14.434~~~  0.011 & 2.630~~~  0.023  & 1.270~~~  0.013 & 0.514~~~  0.010  & 13.684~~~  0.006 & 1.144~~~  0.010  & 1.246~~~  0.006 & 0.708~~~  0.003& 2 \\
SA 101-L6  &   09 57 39 &$-$00 17 54  & 16.901~~~  0.052 & 1.480~~~  0.079  & 0.590~~~  0.070 & 0.384~~~  0.180  & 16.497~~~  0.022 & 0.183~~~  0.031  & 0.711~~~  0.009 & 0.583~~~  0.165& 2 \\
SA 101-207 &   09 57 52 &$-$00 47 36  & 12.626~~~  0.002 & 1.096~~~  0.005  & 0.340~~~  0.003 & 0.102~~~  0.004  & 12.419~~~0.003     & -0.078~~~  0.006 & 0.515~~~  0.004 & 0.320~~~  0.002& 1 \\
SA 101-363 &   09 58 19 &$-$00 25 35  &  9.913~~~  0.006 & 1.090~~~  0.017  & 0.030~~~  0.007 & -0.067~~~  0.010 & 9.874~~~   0.001 & 0.129~~~  0.002  & 0.261~~~  0.001 & 0.151~~~  0.001& 2 \\
PG 1047+003A & 10 50 06 &$-$00 01 09  & 13.821~~~  0.003 & 1.386~~~  0.009  & 0.518~~~  0.004 & 0.212~~~  0.004  & 13.512        & -1.121~~~  0.007 & -0.290~~~  0.005& -0.162~~~  0.003& 1 \\
Ross 106   &   10 50 29 & +56 26 31  & 12.841~~~  0.003 & 1.034~~~  0.008  & 0.429~~~  0.004 & 0.176~~~  0.004  & 12.580        & -0.140        & 0.600        &    0.270 & 1    \\
\end{tabular}  
}  
\end{minipage} 
\end{table*}

\begin{table*}
{\tiny
\begin{tabular}{lccccccccccc}
\hline
Star &    $\alpha$    &    $\delta$   &    $g$          &  $(u-g)$       &  $(g-r)$        &  $(r-i)$        & $V$    &  $(U-B)$      &  $(B-V)$       &  $(R-I)$& Refs. \\
\hline
G163-51    &   11 08 06 &$-$05 13 38  & 13.322$\pm$0.003 & 2.694$\pm$0.004  & 1.326$\pm$0.004 & 1.232$\pm$0.004  & 12.576$\pm$0.005  & 1.228$\pm$0.018  & 1.506$\pm$0.005 & 1.359$\pm$0.004& 1 \\
BD -21 3420 &   11 55 28 &$-$22 23 13  & 10.331~~~0.004 & 1.010~~~0.010  & 0.313~~~0.009 & 0.123~~~0.012  & 10.174~~~0.003   & -0.140~~~0.000  & 0.522~~~0.004 &    0.260& 1    \\
SA 103-626 &   11 56 46 &$-$00 23 12  & 11.982~~~0.003 & 1.073~~~0.009  & 0.234~~~0.004 & 0.057~~~0.004  & 11.836~~~0.003   & -0.057~~~0.003 & 0.413~~~0.003 & 0.274~~~0.002& 1 \\
SA 104-306 &   12 41 04 &$-$00 37 08  & 10.154~~~0.010 & 3.290~~~0.021  & 1.310~~~  0.013 & 0.570~~~0.010  &  9.370~~~0.007  & 1.666~~~0.006  & 1.592~~~0.003 & 0.762~~~0.002& 2 \\
SA 104-423 &   12 41 36 &$-$00 31 04  & 15.841~~~0.085 & 1.320~~~0.127  & 0.250~~~  0.100 & 0.359~~~0.140  & 15.602~~~0.025 & 0.050~~~0.027  & 0.630~~~0.039 & 0.559~~~0.117& 2 \\
SA 104-428 &   12 41 41 &$-$00 26 24  & 13.107~~~0.003 & 2.139~~~0.008  & 0.775~~~0.004 & 0.279~~~0.004  & 12.630~~~0.004& 0.748~~~0.004  & 0.985~~~0.003 & 0.497~~~0.002 & 1\\
SA 104-430 &   12 41 50 &$-$00 25 50  & 14.138~~~0.012 & 1.400~~~0.023  & 0.440~~~0.014 & 0.155~~~0.010  & 13.858~~~0.005 & 0.131~~~0.009  & 0.652~~~ 0.006 & 0.363~~~0.004& 2 \\
SA 104-325 &   12 42 02 &$-$00 41 34  & 15.831~~~0.057 & 1.370~~~0.089  & 0.400~~~0.065 & 0.096~~~0.050  & 15.581~~~0.016 & 0.051~~~0.022  & 0.694~~~ 0.043 & 0.307~~~0.015& 2 \\
SA 104-330 &   12 42 11 &$-$00 40 39  & 15.582~~~0.029 & 1.230~~~0.049  & 0.450~~~0.034 & 0.163~~~0.030  & 15.296~~~0.010 & -0.028~~~0.015 & 0.594~~~0.028 & 0.371~~~0.021& 2 \\
SA 104-440 &   12 42 14 &$-$00 25 02  & 15.317~~~0.019 & 0.960~~~0.031  & 0.300~~~0.023 & 0.107~~~0.030  & 15.114~~~0.008 & -0.227~~~0.010 & 0.440~~~0.008 & 0.317~~~0.023& 2 \\
SA 104-443 &   12 42 20 &$-$00 25 20  & 16.147~~~0.031 & 2.800~~~0.097  & 1.280~~~0.035 & 0.587~~~0.030  & 15.372~~~0.009 & 1.280~~~0.112  & 1.331~~~0.002 & 0.778~~~0.010& 2 \\
SA 104-L2  &   12 42 20 &$-$00 34 22  & 16.301~~~0.068 & 1.160~~~0.105  & 0.400~~~0.079 & 0.113~~~0.100  & 16.048~~~0.019 & -0.172~~~0.045 & 0.650~~~0.027 & 0.323~~~0.076& 2 \\
SA 104-444 &   12 42 21 &$-$00 26 41  & 13.705~~~0.007 & 1.130~~~0.020  & 0.340~~~0.008 & 0.121~~~0.010  & 13.477~~~0.001 & -0.070~~~0.008 & 0.512~~~0.010 & 0.331~~~0.001& 2 \\
SA 104-335 &   12 42 21 &$-$00 33 07  & 11.932~~~0.013 & 1.380~~~0.024  & 0.420~~~0.017 & 0.125~~~0.020  & 11.665~~~0.010 & 0.145~~~0.002  & 0.622~~~0.001 & 0.334~~~0.005& 2 \\
SA 104-334 &   12 42 21 &$-$00 40 27  & 13.720~~~0.010 & 1.140~~~0.020  & 0.360~~~0.012 & 0.121~~~0.010  & 13.484~~~0.005 & -0.067~~~0.004 & 0.518~~~0.003 & 0.331~~~0.003& 2 \\
SA 104-336 &   12 42 25 &$-$00 39 56  & 14.778~~~0.015 & 1.810~~~0.027  & 0.620~~~0.018 & 0.197~~~0.020  & 14.404~~~0.007 & 0.495~~~0.012  & 0.830~~~0.007 & 0.403~~~0.010& 2 \\
SA 104-338 &   12 42 30 &$-$00 38 31  & 16.327~~~0.022 & 1.190~~~0.047  & 0.410~~~0.026 & 0.164~~~0.030  & 16.059~~~0.008 & -0.082~~~0.037 & 0.591~~~0.022 & 0.372~~~0.021& 2 \\ 
SA 104-244 &   12 42 34 &$-$00 45 48  & 16.299~~~0.015 & 1.130~~~0.027  & 0.390~~~0.019 & 0.286~~~0.030  & 16.011~~~0.009 & -0.152~~~0.011 & 0.590~~~0.006 & 0.489~~~0.027& 2 \\
SA 104-455 &   12 42 52 &$-$00 24 16 & 15.380~~~0.014 & 1.220~~~0.029  & 0.430~~~0.017 & 0.148~~~0.020  & 15.105~~~0.007 & -0.024~~~0.012 & 0.581~~~0.014 & 0.357~~~0.009& 2 \\
SA 104-457 &   12 42 54 &$-$00 28 48  & 16.463~~~0.042 & 1.770~~~0.086  & 0.660~~~0.048 & 0.287~~~0.050  & 16.048~~~0.011 & 0.522~~~0.077  & 0.753~~~0.019 & 0.490~~~0.037& 2  \\
SA 104-456 &   12 42 54 &$-$00 32 06  & 12.629~~~0.010 & 1.380~~~0.021  & 0.420~~~0.012 & 0.127~~~0.010  & 12.362~~~0.003 & 0.135~~~0.005  & 0.622~~~0.003 & 0.337~~~0.007& 2 \\
SA 104-460 &   12 43 03 &$-$00 28 17  & 13.636~~~0.009 & 2.730~~~0.019  & 1.280~~~0.011 & 0.498~~~0.010  & 12.886~~~0.004 & 1.243~~~0.007  & 1.287~~~0.003 & 0.693~~~0.002& 2 \\
SA 104-461 &   12 43 06 &$-$00 32 15  &  9.902~~~0.006 & 1.140~~~0.016  & 0.300~~~0.007 & 0.078~~~0.010  &  9.705~~~0.001 & -0.030~~~0.001 & 0.476~~~0.001 & 0.290~~~0.001& 2  \\
SA 104-350 &   12 43 14 &$-$00 33 18  & 13.928~~~0.010 & 1.440~~~0.021  & 0.470~~~0.012 & 0.144~~~0.010  & 13.634~~~0.004 & 0.165~~~0.006  & 0.673~~~0.003 & 0.353~~~0.005& 2 \\
SA 104-470 &   12 43 22 &$-$00 29 52  & 14.528~~~0.031 & 1.430~~~0.049  & 0.310~~~0.036 & 0.147~~~0.030  & 14.310~~~0.012 & 0.101~~~0.020  & 0.732~~~0.009 & 0.356~~~0.003& 2 \\
SA 104-364 &   12 43 46 &$-$00 34 32 & 16.044~~~0.026 & 1.160~~~0.069  & 0.340~~~0.033 & 0.190~~~0.050  & 15.799~~~0.016 & -0.131~~~0.056 & 0.601~~~0.049 & 0.397~~~0.041& 2 \\
SA 104-479 &   12 43 55 &$-$00 32 48 & 16.681~~~0.035 & 2.280~~~0.311  & 0.990~~~0.049 & 0.408~~~0.050  & 16.087~~~0.034 & 0.673~~~0.400  & 1.271~~~0.028 & 0.607~~~0.004& 2 \\ 
SA 104-367 &   12 43 59 &$-$00 33 31 & 16.123~~~0.046 & 1.190~~~0.069  & 0.470~~~0.055 & 0.084~~~0.040  & 15.844~~~0.025 & -0.126~~~0.016 & 0.639~~~0.021 & 0.296~~~0.004& 2 \\
SA 104-484 &   12 44 21 &$-$00 30 54  & 14.846~~~0.015 & 2.140~~~0.031  & 0.720~~~0.018 & 0.283~~~0.020  & 14.406~~~0.007 & 0.732~~~0.013  & 1.024~~~0.018 & 0.486~~~0.006& 2 \\
SA 104-485 &   12 44 24 &$-$00 30 14  & 15.426~~~0.013 & 1.820~~~0.038  & 0.650~~~0.018 & 0.284~~~0.020  & 15.017~~~0.011 & 0.493~~~0.023  & 0.838~~~0.034 & 0.488~~~0.011& 2 \\
SA 104-490 &   12 44 34 &$-$00 25 48  & 12.799~~~0.008 & 1.240~~~0.019  & 0.350~~~0.009 & 0.101~~~0.010  & 12.572~~~0.003 & 0.048~~~0.005  & 0.535~~~0.003 & 0.312~~~0.002& 2 \\
SA 104-598 &   12 45 17 &$-$00 16 38  & 12.069~~~0.006 & 2.454~~~0.015  & 1.007~~~0.006 & 0.338~~~0.006  & 11.479~~~0.002 & 1.050~~~0.003  & 1.106~~~0.001 & 0.546~~~0.001& 1 \\
LTT 5137   &   13 20 24 &$-$03 01 42 & 11.641~~~0.004 & 1.107~~~0.011  & 0.452~~~0.006 & 0.181~~~0.005  & 11.040         & -0.100         & 0.580        & 0.270& 1  \\
PG 1323-086D & 13 26 05 &$-$08 50 34 & 12.317~~~0.001 & 1.218~~~0.003  & 0.392~~~0.001 & 0.132~~~0.001  & 12.080~~~0.002 & 0.005~~~0.004  & 0.587~~~0.002 & 0.335~~~0.003& 1 \\
G14-55     &   13 28 21 &$-$02 21 26 & 12.006~~~0.003 & 2.765~~~0.007  & 1.334~~~0.004 & 1.326~~~0.003  & 11.336~~~0.006 & 1.157~~~0.008  & 1.491~~~0.005 & 1.388~~~0.001& 1 \\
BD +30 2428B &  13 37 14 & +30 05 14  & 10.915~~~0.003 & 1.836~~~0.006  & 0.676~~~0.004 & 0.223~~~0.004  & 10.517         & 0.557         & 0.842       &   0.265& 1 \\
SA 105-815 &   13 40 03 &$-$00 02 15 & 11.580~~~0.002 & 0.881~~~0.005  & 0.219~~~0.003 & 0.077~~~0.004  & 11.453~~~0.002 & -0.237~~~0.003 & 0.385~~~0.003 & 0.291~~~0.001& 1 \\
GCRV 8758  &   15 07 41 & +32 24 37  & 11.444~~~0.002 & 1.260~~~0.004  & 0.491~~~0.003 & 0.176~~~0.003  & 11.127~~~0.006 & 0.130~~~0.018  & 0.677~~~0.008 &     0.300& 1     \\
PG 1528+062B & 15 30 40 & +06 01 13  & 12.240~~~0.001 & 1.242~~~0.004  & 0.414~~~0.001 & 0.143~~~0.001  & 11.989~~~0.000 & 0.005~~~0.000  & 0.593~~~0.000 &   0.344~~~0.000& 1  \\
G15-24     &   15 30 41 & +08 23 41  & 11.682~~~0.002 & 1.042~~~0.004  & 0.407~~~0.003 & 0.151~~~0.003  & 11.435         & -0.140        & 0.570        &    0.360& 1    \\
SA 107-568 &   15 37 53 &$-$00 17 15  & 13.617~~~0.007 & 2.340~~~0.017  & 0.920~~~0.008 & 0.397~~~0.010  & 13.054~~~0.002 & 0.862~~~0.004  & 1.149~~~0.003 & 0.595~~~0.003& 2 \\
SA107-1006 &   15 38 33 & +00 14 19  & 12.055~~~0.006 & 1.546~~~0.007  & 0.581~~~0.003 & 0.204~~~0.003  & 11.712~~~0.001 & 0.279~~~0.003  & 0.766~~~0.001 & 0.421~~~0.001& 1 \\
SA 107-720 &   15 38 37 &$-$00 02 23  & 13.408~~~0.022 & 1.320~~~0.036  & 0.460~~~0.025 & 0.146~~~0.020  & 13.121~~~0.006 & 0.088~~~0.011  & 0.599~~~0.003 & 0.355~~~0.003& 2 \\
SA 107-456 &   15 38 43 &$-$00 19 47  & 13.377~~~0.007 & 1.950~~~0.017  & 0.760~~~0.008 & 0.274~~~0.010  & 12.919~~~0.002 & 0.589~~~0.004  & 0.921~~~0.003 & 0.478~~~0.002& 2 \\
SA107-351  &   15 38 46 &$-$00 32 04 & 12.563~~~0.007 & 1.195~~~0.004  & 0.390~~~0.003 & 0.142~~~0.003  & 12.342~~~0.002 & -0.005~~~0.004 & 0.562~~~0.002 & 0.358~~~0.002& 1 \\
SA 107-457 &   15 38 47 &$-$00 20 15  & 15.329~~~0.009 & 1.670~~~0.020  & 0.680~~~0.010 & 0.265~~~0.010  & 14.910~~~0.002 & 0.350~~~0.010  & 0.792~~~0.004 & 0.469~~~0.006& 2 \\
SA 107-592 &   15 38 50 &$-$00 17 07  & 12.496~~~0.007 & 2.860~~~0.020  & 1.080~~~0.009 & 0.450~~~0.010  & 11.847~~~0.001 & 1.380~~~0.013  & 1.318~~~0.005 & 0.647~~~0.002& 2 \\
SA 107-458 &   15 38 50 &$-$00 24 26  & 12.277~~~0.009 & 2.640~~~0.020  & 1.000~~~0.011 & 0.403~~~0.010  & 11.676~~~0.004 & 1.189~~~0.005  & 1.214~~~0.006 & 0.602~~~0.001& 2 \\
SA 107-459 &   15 38 51 &$-$00 22 34  & 12.741~~~0.008 & 1.810~~~0.022  & 0.740~~~0.009 & 0.315~~~0.010  & 12.284~~~0.003 & 0.427~~~0.010  & 0.900~~~0.011 & 0.517~~~0.001& 2 \\
SA 107-212 &   15 38 56 &$-$00 45 30  & 13.709~~~0.017 & 1.420~~~0.033  & 0.510~~~0.020 & 0.205~~~0.020  & 13.383~~~0.006 & 0.135~~~0.020  & 0.683~~~0.011 & 0.411~~~0.008& 2 \\ 
SA 107-357 &   15 39 05 &$-$00 39 12  & 14.757~~~0.027 & 1.330~~~0.041  & 0.530~~~0.030 & 0.215~~~0.030  & 14.418~~~0.000 & 0.025~~~0.006  & 0.675~~~0.004 & 0.421~~~0.023& 2 \\
SA 107-359 &   15 39 09 &$-$00 35 37  & 13.094~~~0.010 & 1.150~~~0.022  & 0.470~~~0.012 & 0.173~~~0.010  & 12.797~~~0.005 & -0.124~~~0.012 & 0.580~~~0.007 & 0.381~~~0.003& 2 \\ 
SA 107-599 &   15 39 10 &$-$00 14 26  & 15.033~~~0.015 & 1.520~~~0.027  & 0.570~~~0.019 & 0.232~~~0.020  & 14.675~~~0.008 & 0.243~~~0.006  & 0.698~~~0.012 & 0.438~~~0.012& 2 \\
SA 107-600 &   15 39 10 &$-$00 15 47  & 15.142~~~0.027 & 1.220~~~0.046  & 0.390~~~0.034 & 0.152~~~0.030  & 14.884~~~0.017 & 0.049~~~0.023  & 0.503~~~0.013 & 0.361~~~0.015& 2 \\    
SA 107-602 &   15 39 19 &$-$00 15 30  & 12.594~~~0.007 & 2.000~~~0.018  & 0.770~~~0.009 & 0.330~~~0.010  & 12.116~~~0.003 & 0.585~~~0.007  & 0.991~~~0.003 & 0.531~~~0.002& 2 \\   
SA 107-611 &   15 39 35 &$-$00 12 35  & 14.765~~~0.021 & 1.830~~~0.036  & 0.730~~~0.024 & 0.242~~~0.020  & 14.329~~~0.007 & 0.455~~~0.017  & 0.890~~~0.009 & 0.447~~~0.011& 2 \\
SA 107-612 &   15 39 35 &$-$00 15 28   & 14.739~~~0.011 & 1.710~~~0.021  & 0.790~~~0.013 & 0.329~~~0.010  & 14.256~~~0.006 & 0.296~~~0.004  & 0.896~~~0.001 & 0.530~~~0.001& 2  \\
SA 107-614 &   15 39 41 &$-$00 13 10  & 14.205~~~0.012 & 1.300~~~0.028  & 0.430~~~0.015 & 0.162~~~0.020  & 13.926~~~0.007 & 0.033~~~0.015  & 0.622~~~0.017 & 0.370~~~0.013& 2  \\
SA 107-626 &   15 40 06 &$-$00 17 28  & 13.993~~~0.009 & 2.120~~~0.022  & 0.880~~~0.011 & 0.325~~~0.010  & 13.468~~~0.004 & 0.728~~~0.012  & 1.000~~~0.008 & 0.527~~~0.002& 2 \\
SA 107-627 &   15 40 07 &$-$00 17 22  & 13.739~~~0.009 & 1.570~~~0.020  & 0.630~~~0.011 & 0.249~~~0.010  & 13.349~~~0.006 & 0.226~~~0.005  & 0.779~~~0.005 & 0.454~~~0.003& 2 \\
SA 107-484 &   15 40 17 &$-$00 21 10  & 11.904~~~0.008 & 2.730~~~0.020  & 1.000~~~0.010 & 0.377~~~0.010  & 11.311~~~0.004 & 1.291~~~0.011  & 1.237~~~0.002 & 0.577~~~0.001& 2 \\
SA 107-636 &   15 40 40 &$-$00 14 52  & 15.237~~~0.025 & 1.460~~~0.041  & 0.560~~~0.031 & 0.261~~~0.050  & 14.873~~~0.013 & 0.121~~~0.018  & 0.751~~~0.012 & 0.465~~~0.042& 2 \\
SA 107-639 &   15 40 45 &$-$00 17 10  & 14.517~~~0.028 & 1.270~~~0.046  & 0.500~~~0.034 & 0.197~~~0.030  & 14.197~~~0.016 & -0.026~~~0.020  & 0.640~~~0.009 & 0.404~~~0.009& 2 \\
SA 107-640 &   15 40 49 &$-$00 16 47  & 15.492~~~0.047 & 1.450~~~0.072  & 0.710~~~0.056 & 0.304~~~0.050  & 15.050~~~0.025 & 0.092~~~0.032  & 0.755~~~0.010 & 0.506~~~0.016& 2 \\
SA 108-551 &   16 37 48 &$-$00 33 05  & 10.681~~~0.003 & 1.290~~~0.006  & -0.055~~~0.004& -0.102~~~0.003& 10.703~~~0.001& 0.178~~~0.002  & 0.179~~~0.001  & 0.110~~~0.001& 1 \\
Wolf 629   &   16 55 25 &$-$08 19 21  & 12.595~~~0.004 & 2.960~~~0.021  & 1.422~~~0.004 & 1.458~~~0.004 & 11.759~~~0.004& 1.256~~~0.003   & 1.677~~~0.003 & 1.525~~~0.001& 1 \\
BD +18 3407 &   17 35 20 & +18 53 01  & 10.399~~~0.001  & 1.684~~~0.003  & 0.633~~~0.001 & 0.227~~~0.001  & 10.050        & 0.420           & 0.790        &    0.430& 1    \\
SA 109-71  &   17 44 07 &$-$00 24 58  & 11.581~~~0.002 & 1.298~~~0.004  & 0.113~~~0.002 & 0.011~~~0.001 & 11.493~~~0.001& 0.153~~~0.002   & 0.323~~~0.001 & 0.223~~~0.001& 1 \\
SA 109-381 &   17 44 12 &$-$00 20 33  & 12.062~~~0.002 & 1.476~~~0.006  & 0.548~~~0.003 & 0.223~~~0.002 & 11.730~~~0.001& 0.225~~~0.002   & 0.704~~~0.002 & 0.435~~~0.001& 1 \\
SA 109-537 &   17 45 42 &$-$00 21 36  & 10.653~~~0.003 & 1.453~~~0.008  & 0.453~~~0.004 & 0.191~~~0.004 & 10.353~~~0.003& 0.227~~~0.002   & 0.609~~~0.001 & 0.392~~~0.001& 1 \\
SA 110-232 &   18 40 52 & +00 01 55  & 12.840~~~0.002 & 1.389~~~0.005  & 0.552~~~0.002 & 0.237~~~0.002  & 12.516~~~0.003 & 0.147~~~0.005  & 0.729~~~0.003 & 0.450~~~0.002& 1 \\
SA 110-230 &   18 40 52 & +00 02 23  & 14.843~~~0.008 & 2.180~~~0.021  & 0.920~~~0.009 & 0.398~~~0.010  & 14.281~~~0.003 & 0.728~~~0.012  & 1.084~~~0.005 & 0.596~~~0.004& 2 \\
SA 110-340 &   18 41 29 & +00 15 23  & 10.092~~~0.006 & 1.130~~~0.017  & 0.070~~~0.007 & -0.034~~~0.010 & 10.025~~~0.001 & 0.127~~~0.002  & 0.303~~~0.001 & 0.182~~~0.001& 2 \\
SA 110-355 &   18 42 19 & +00 08 24  & 12.563~~~0.008 & 1.970~~~0.019  & 0.980~~~0.009 & 0.534~~~0.010  & 11.944~~~0.003 & 0.504~~~0.005  & 1.023~~~0.000 & 0.727~~~0.002& 2 \\
SA 110-360 &   18 42 41 & +00 09 11  & 15.289~~~0.021 & 2.130~~~0.043  & 1.090~~~0.025 & 0.523~~~0.020  & 14.618~~~0.009 & 0.539~~~0.029  & 1.197~~~0.019 & 0.717~~~0.003& 2 \\
SA 110-361 &   18 42 45 & +00 08 05  & 12.699~~~0.007 & 1.310~~~0.017  & 0.430~~~0.008 & 0.139~~~0.010  & 12.425~~~0.002 & 0.035~~~0.003  & 0.632~~~0.002 & 0.348~~~0.002& 2 \\
SA 110-266 &   18 42 49 & +00 05 00  & 12.501~~~0.007 & 1.790~~~0.017  & 0.760~~~0.008 & 0.377~~~0.010  & 12.018~~~0.001 & 0.4110.004  & 0.889~~~0.003 & 0.577~~~0.002& 2 \\
SA 110-364 &   18 42 52 & +00 08 00  & 14.238~~~0.007 & 2.500~~~0.019  & 1.060~~~0.008 & 0.386~~~0.010  & 13.615~~~0.002 & 1.095~~~0.009  & 1.133~~~0.007 & 0.585~~~0.003& 2 \\
SA 110-496 &   18 42 59 & +00 31 00  & 13.572~~~0.008 & 2.160~~~0.025  & 0.890~~~0.009 & 0.486~~~0.010  & 13.004~~~0.003 & 0.737~~~0.023  & 1.040~~~0.006 & 0.681~~~0.002& 2 \\
SA 110-497 &   18 43 02 & +00 31 00  & 14.743~~~0.016 & 1.890~~~0.029  & 0.890~~~0.018 & 0.398~~~0.010  & 14.196~~~0.005 & 0.38~~~0.014   & 1.052~~~0.006 & 0.597~~~0.004& 2 \\ 
\end{tabular}  
}  
\end{table*}

\begin{table*}
{\tiny
\begin{tabular}{lccccccccccc}
\hline
Star &    $\alpha$    &    $\delta$   &    $g$          &  $(u-g)$       &  $(g-r)$        &  $(r-i)$        & $V$    &  $(U-B)$      &  $(B-V)$       &  $(R-I)$& Refs. \\
\hline
SA 110-499 &   18 43 08 & +00 28 01  & 12.298$\pm$0.008 & 2.040$\pm$0.019  & 0.880$\pm$0.009 & 0.479$\pm$0.010  & 11.737$\pm$0.003 & 0.639$\pm$0.007  & 0.987$\pm$0.003 & 0.674$\pm$0.002& 2 \\
SA 110-503 &   18 43 12 & +00 29 43  & 12.081~~~0.002 & 1.796~~~0.005  & 0.456~~~0.003 & 0.222~~~0.004  & 11.773~~~0.003 & 0.506~~~0.005  & 0.671~~~0.002 & 0.436~~~0.002& 1 \\ 
SA 110-504 &   18 43 11 & +00 30 05  & 14.756~~~0.010 & 2.770~~~0.029  & 1.250~~~0.011 & 0.488~~~0.010  & 14.022~~~0.001 & 1.323~~~0.027  & 1.248~~~0.006 & 0.683~~~0.006& 2 \\ 
SA 110-506 &   18 43 19 & +00 30 27  & 11.554~~~0.006 & 1.280~~~0.017  & 0.380~~~0.007 & 0.101~~~0.010  & 11.312~~~0.002 & 0.059~~~0.006  & 0.568~~~0.002 & 0.312~~~0.004& 2 \\
SA 110-507 &   18 43 20 & +00 29 26  & 13.006~~~0.012 & 2.310~~~0.022  & 0.940~~~0.014 & 0.380~~~0.010  & 12.44~~~0.005 & 0.830~~~0.006   & 1.141~~~0.006 &   0.579~~~0.000& 2 \\
SA 110-290 &   18 43 22 &$-$00 01 00  & 12.238~~~0.008 & 1.490~~~0.019  & 0.540~~~0.009 & 0.212~~~0.010  & 11.898~~~0.003 & 0.196~~~0.005  & 0.708~~~0.001 & 0.418~~~0.003& 2 \\
SA 110-441 &   18 43 34 & +00 20 00  & 11.377~~~0.028 & 1.310~~~0.043  & 0.370~~~0.033 & 0.139~~~0.020  & 11.121~~~0.002 & 0.112~~~0.002  & 0.555~~~0.002 & 0.336~~~0.001& 2 \\
SA 110-450 &   18 43 52 & +00 23 00  & 12.077~~~0.026 & 2.120~~~0.061  & 0.770~~~0.030 & 0.430~~~0.020  & 11.585~~~0.002 & 0.691~~~0.003  & 0.944~~~0.002 & 0.625~~~0.001& 2 \\
SA 110-319 &   18 43 55 & +00 02 00  & 12.551~~~0.015 & 2.620~~~0.026  & 1.140~~~0.019 & 0.506~~~0.020  & 11.861~~~0.011 & 1.076~~~0.003  & 1.309~~~0.003 & 0.700~~~0.002& 2 \\
GJ 745A    &   19 07 06 & +20 53 17  & 11.620~~~0.002 & 2.795~~~0.004  & 1.432~~~0.002 & 1.041~~~0.002  & 10.780         & 1.165          & 1.470        &   1.305& 1     \\
GJ 745B    &   19 07 13 & +20 52 37  & 11.604~~~0.002 & 2.787~~~0.006  & 1.424~~~0.002 & 1.044~~~0.002  & 10.770         & 1.150          & 1.470        &    1.230& 1    \\
SA 111-775 &   19 37 16 & +00 12 05  & 11.655~~~0.004 & 3.707~~~0.019  & 1.532~~~0.005 & 0.694~~~0.004  & 10.744~~~0.002 & 2.029~~~0.006  & 1.738~~~0.001 & 0.896~~~0.001& 1 \\ 
SA 111-1925 &  19 37 29 & +00 25 03  & 12.525~~~0.003 & 1.417~~~0.011  & 0.185~~~0.004 & 0.062~~~0.005  & 12.388~~~0.002 & 0.262~~~0.004  & 0.395~~~0.001 & 0.253~~~0.001& 1 \\
SA 112-223 &   20 42 15 & +00 09 01  & 11.590~~~0.003 & 1.161~~~0.006  & 0.259~~~0.004 & 0.063~~~0.003  & 11.424~~~0.001 & 0.010~~~0.002  & 0.454~~~0.001 & 0.274~~~0.001& 1 \\
SA 112-250 &   20 42 26 & +00 07 45  & 12.303~~~0.003 & 1.173~~~0.008  & 0.344~~~0.004 & 0.117~~~0.003  & 12.095~~~0.002 & -0.025~~~0.004 & 0.532~~~0.002 & 0.323~~~0.002& 1 \\
BD +62 1916 &   21 15 06 & +62 50 28  &  9.873~~~0.002 & 1.567~~~0.004  & 0.542~~~0.002 & 0.171~~~0.001  &  9.535~~~0.040 & 0.343~~~0.080  & 0.751~~~0.060 &    0.310& 1    \\
SA 113-440 &   21 40 34 & +00 41 48  & 12.072~~~0.006 & 1.410~~~0.017  & 0.430~~~0.007 & 0.141~~~0.010  & 11.796~~~0.001 & 0.167~~~0.004  & 0.637~~~0.001 & 0.350~~~0.003& 2 \\
SA 113-221 &   21 40 37 & +00 21 05  & 12.543~~~0.007 & 2.260~~~0.017  & 0.780~~~0.008 & 0.287~~~0.010  & 12.071~~~0.002 & 0.874~~~0.006  & 1.031~~~0.002 & 0.490~~~0.001& 2 \\
SA 113-L1  &   21 40 47 & +00 28 37  & 16.335~~~0.013 & 2.730~~~0.150  & 1.380~~~0.015 & 0.530~~~0.010  & 15.530~~~0.004 & 1.180~~~0.187  & 1.343~~~0.051 & 0.723~~~0.005& 2 \\
SA 113-337 &   21 40 50 & +00 28 00  & 14.486~~~0.021 & 1.170~~~0.033  & 0.410~~~0.026 & 0.121~~~0.020  & 14.225~~~0.013 & -0.025~~~0.009 & 0.519~~~0.006 & 0.331~~~0.006& 2 \\
SA 113-339 &   21 40 56 & +00 28 00  & 12.484~~~0.007 & 1.157~~~0.004  & 0.384~~~0.002 & 0.127~~~0.001  & 12.250~~~0.002  & -0.034~~~0.004 & 0.568~~~0.002 & 0.347~~~0.002& 1 \\
SA 113-233 &   21 40 59 & +00 22 05  & 12.645~~~0.006 & 1.290~~~0.017  & 0.390~~~0.007 & 0.112~~~0.010  & 12.398~~~0.002 & 0.096~~~0.001  & 0.549~~~0.001 & 0.322~~~0.001& 2 \\
SA 113-342 &   21 41 00 & +00 27 40  & 11.345~~~0.007 & 2.110~~~0.018  & 0.760~~~0.008 & 0.310~~~0.010  & 10.878~~~0.002 & 0.696~~~0.006  & 1.015~~~0.002 & 0.513~~~0.002& 2 \\
SA 113-239 &   21 41 07 & +00 22 36  & 13.269~~~0.009 & 1.230~~~0.019  & 0.350~~~0.011 & 0.117~~~0.010  & 13.038~~~0.006 & 0.051~~~0.004  & 0.516~~~0.003 & 0.327~~~0.004& 2 \\
SA 113-241 &   21 41 09 & +00 25 51  & 15.201~~~0.007 & 2.940~~~0.017  & 1.430~~~0.008 & 0.607~~~0.010  & 14.352~~~0.002 & 1.452~~~0.006  & 1.344~~~0.003 & 0.797~~~0.003& 2 \\
SA 113-245 &   21 41 13 & +00 21 54  & 15.962~~~0.052 & 1.360~~~0.076  & 0.500~~~0.058 & 0.108~~~0.040  & 15.665~~~0.009 & 0.112~~~0.002  & 0.628~~~0.010 & 0.318~~~0.011& 2 \\
SA 113-459 &   21 41 15 & +00 43 05  & 12.343~~~0.015 & 1.190~~~0.026  & 0.330~~~0.018 & 0.103~~~0.020  & 12.125~~~0.007 & -0.018~~~0.007 & 0.535~~~0.006 & 0.313~~~0.006& 2 \\
SA 113-250 &   21 41 25 & +00 20 45  & 13.381~~~0.010 & 1.180~~~0.023  & 0.340~~~0.012 & 0.105~~~0.010  & 13.160~~~0.003 & -0.003~~~0.017 & 0.505~~~0.003 & 0.316~~~0.005& 2 \\
SA 113-466 &   21 41 27 & +00 40 16  & 10.168~~~0.006 & 1.140~~~0.003  & 0.265~~~0.002 & 0.074~~~0.002  & 10.004~~~0.001 & -0.001~~~0.002 & 0.454~~~0.001 & 0.282~~~0.001& 1 \\
SA 113-259 &   21 41 45 & +00 17 42  & 12.314~~~0.011 & 2.780~~~0.051  & 0.940~~~0.014 & 0.346~~~0.010  & 11.742~~~0.001 & 1.221~~~0.003  & 1.194~~~0.002 & 0.543~~~0.001& 2 \\
SA 113-260 &   21 41 48 & +00 23 53  & 12.603~~~0.014 & 1.229~~~0.006  & 0.324~~~0.003 & 0.081~~~0.003  & 12.406~~~0.004 & 0.069~~~0.003  & 0.514~~~0.002 & 0.298~~~0.002& 1 \\
SA 113-475 &   21 41 51 & +00 39 24  & 10.805~~~0.006 & 2.250~~~0.017  & 0.820~~~0.007 & 0.325~~~0.010  & 10.306~~~0.001 & 0.844~~~0.003  & 1.058~~~0.001 & 0.527~~~0.001& 2 \\
SA 113-263 &   21 41 53 & +00 25 39  & 15.575~~~0.014 & 1.070~~~0.031  & 0.120~~~0.018 & -0.008~~~0.020 & 15.481~~~0.010 & 0.074~~~0.025  & 0.280~~~0.008 & 0.207~~~0.012& 2 \\
SA 113-366 &   21 41 54 & +00 29 25  & 14.097~~~0.025 & 2.320~~~0.038  & 0.920~~~0.028 & 0.389~~~0.020  & 13.537~~~0.001 & 0.896~~~0.005  & 1.096~~~0.005 & 0.588~~~0.002& 2 \\
SA 113-265 &   21 41 54 & +00 18 05  & 15.262~~~0.039 & 1.360~~~0.059  & 0.520~~~0.046 & 0.188~~~0.040  & 14.934~~~0.018 & 0.101~~~0.016  & 0.639~~~0.012 & 0.395~~~0.017& 2 \\
SA 113-268 &   21 41 57 & +00 19 57  & 15.585~~~0.018 & 1.230~~~0.035  & 0.470~~~0.021 & 0.201~~~0.030  & 15.281~~~0.003 & -0.018~~~0.023 & 0.589~~~0.009 & 0.407~~~0.021& 2 \\
SA 113-34  &   21 41 59 & +00 01 08  & 15.398~~~0.023 & 1.130~~~0.039  & 0.330~~~0.026 & 0.137~~~0.020  & 15.173~~~0.001 & -0.054~~~0.021 & 0.484~~~0.004 & 0.346~~~0.013& 2 \\
SA 113-372 &   21 42 02 & +00 28 41  & 13.989~~~0.012 & 1.370~~~0.024  & 0.490~~~0.014 & 0.162~~~0.010  & 13.681~~~0.004 & 0.080~~~0.014  & 0.670~~~0.003 & 0.370~~~0.010& 2 \\
SA 113-149 &   21 42 06 & +00 09 27  & 13.767~~~0.014 & 1.310~~~0.028  & 0.470~~~0.017 & 0.178~~~0.020  & 13.469~~~0.008 & 0.043~~~0.010  & 0.621~~~0.015 & 0.386~~~0.008& 2 \\
SA 113-153 &   21 42 09 & +00 15 06  & 14.860~~~0.108 & 1.590~~~0.161  & 0.620~~~0.151 & 0.236~~~0.150  & 14.776~~~0.105 & 0.285~~~0.050  & 0.745~~~0.038 & 0.441~~~0.014& 2 \\
SA 113-272 &   21 42 20 & +00 21 01  & 14.183~~~0.007 & 1.330~~~0.026  & 0.450~~~0.009 & 0.130~~~0.020  & 13.904~~~0.001 & 0.067~~~0.025  & 0.633~~~0.001 & 0.340~~~0.013& 2  \\
SA 113-156 &   21 42 22 & +00 12 11  & 11.439~~~0.011 & 1.160~~~0.022  & 0.320~~~0.013 & 0.104~~~0.010  & 11.224~~~0.004 & -0.057~~~0.006 & 0.526~~~0.004 & 0.314~~~0.001& 2 \\
SA 113-158 &   21 42 22 & +00 14 11  & 13.436~~~0.009 & 1.540~~~0.019  & 0.520~~~0.011 & 0.166~~~0.010  & 13.116~~~0.003 & 0.247~~~0.004  & 0.723~~~0.002 & 0.374~~~0.005& 2 \\
SA 113-491 &   21 42 25 & +00 43 57  & 14.728~~~0.019 & 1.620~~~0.034  & 0.570~~~0.021 & 0.214~~~0.020  & 14.373~~~0.003 & 0.306~~~0.019  & 0.764~~~0.002 & 0.420~~~0.010& 2  \\
SA 113-492 &   21 42 28 & +00 38 25  & 12.429~~~0.009 & 1.220~~~0.020  & 0.400~~~0.010 & 0.131~~~0.010  & 12.174~~~0.003 & 0.005~~~0.006  & 0.553~~~0.006 & 0.341~~~0.003& 2 \\
SA 113-493 &   21 42 29 & +00 38 15  & 12.111~~~0.009 & 1.700~~~0.020  & 0.560~~~0.011 & 0.186~~~0.010  & 11.767~~~0.004 & 0.392~~~0.006  & 0.786~~~0.004 & 0.393~~~0.002& 2 \\ 
SA 113-495 &   21 42 30 & +00 38 10  & 12.878~~~0.009 & 1.930~~~0.020  & 0.710~~~0.010 & 0.295~~~0.010  & 12.437~~~0.002 & 0.530~~~0.005  & 0.947~~~0.003 & 0.497~~~0.004& 2 \\
SA 113-163 &   21 42 35 & +00 16 48  & 14.832~~~0.014 & 1.380~~~0.025  & 0.470~~~0.016 & 0.146~~~0.010  & 14.540~~~0.004 & 0.106~~~0.009  & 0.658~~~0.005 & 0.355~~~0.008& 2 \\ 
SA 113-165 &   21 42 38 & +00 15 35  & 15.917~~~0.046 & 1.260~~~0.071  & 0.420~~~0.052 & 0.185~~~0.060  & 15.639~~~0.001 & 0.003~~~0.030  & 0.601~~~0.006 & 0.392~~~0.043& 2 \\
SA 113-281 &   21 42 39 & -00 18 59  & 15.511~~~0.014 & 1.180~~~0.034  & 0.400~~~0.018 & 0.150~~~0.040  & 15.247~~~0.008 & -0.026~~~0.031 & 0.529~~~0.001 & 0.359~~~0.037& 2 \\
SA 113-167 &   21 42 41 & +00 16 11  & 15.113~~~0.016 & 1.230~~~0.033  & 0.410~~~0.018 & 0.168~~~0.020  & 14.841~~~0.001 & -0.034~~~0.024 & 0.597~~~0.010 & 0.376~~~0.014& 2 \\
SA 113-177 &   21 42 56 & +00 14 46  & 13.938~~~0.012 & 1.640~~~0.022  & 0.610~~~0.014 & 0.231~~~0.010  & 13.560~~~0.005 & 0.318~~~0.006  & 0.789~~~0.004 & 0.436~~~0.004& 2 \\
SA 113-182 &   21 43 08 & +00 14 52  & 14.697~~~0.018 & 1.350~~~0.030  & 0.510~~~0.023 & 0.216~~~0.020  & 14.370~~~0.014 & 0.065~~~0.009  & 0.659~~~0.009 & 0.422~~~0.009& 2 \\
SA 113-187 &   21 43 20 & +00 16 51  & 15.640~~~0.017 & 2.350~~~0.038  & 0.950~~~0.020 & 0.334~~~0.010  & 15.080~~~0.006 & 0.969~~~0.030  & 1.063~~~0.016 & 0.535~~~0.004& 2 \\
SA 113-189 &   21 43 27 & +00 17 24  & 16.063~~~0.028 & 2.390~~~0.070  & 1.090~~~0.033 & 0.407~~~0.030 & 15.421~~~0.014 & 0.958~~~0.071  & 1.118~~~0.013 & 0.605~~~0.007& 2 \\
SA 113-307 &   21 43 30 & +00 18 06  & 14.786~~~0.010 & 2.360~~~0.029  & 0.930~~~0.013 & 0.416~~~0.030  & 14.214~~~0.005 & 0.911~~~0.004  & 1.128~~~0.027 & 0.614~~~0.022& 2 \\
SA 113-191 &   21 43 33 & +00 15 57  & 12.735~~~0.009 & 1.580~~~0.019  & 0.640~~~0.011 & 0.262~~~0.010  & 12.337~~~0.004 & 0.223~~~0.005  & 0.799~~~0.002 & 0.466~~~0.002& 2 \\
SA 113-195 &   21 43 41 & +00 17 24  & 14.031~~~0.010 & 1.510~~~0.021  & 0.540~~~0.012 & 0.207~~~0.010  & 13.692~~~0.005 & 0.201~~~0.010  & 0.730~~~0.006 & 0.413~~~0.004& 2 \\ 
BD -11 5781 &  22 13 11 &$-$11 10 38  &  9.909~~~0.004 & 1.955~~~0.009  & 0.730~~~0.006 & 0.236~~~0.005 &  9.424~~~0.005 & 0.653~~~0.009  & 0.917~~~0.007 &   0.449& 1     \\
SA 114-531 &   22 40 37 & +00 51 58  & 12.421~~~0.002 & 1.418~~~0.010  & 0.542~~~0.003 & 0.187~~~0.003 & 12.094~~~0.001 & 0.186~~~0.003  & 0.733~~~0.002 & 0.403~~~0.001& 1 \\
SA 114-654 &   22 41 26 & +01 10 13  & 12.116~~~0.001 & 1.403~~~0.004  & 0.447~~~0.001 & 0.137~~~0.001  & 11.833~~~0.001 & 0.178~~~0.006  & 0.656~~~0.004 & 0.341~~~0.003& 1 \\
SA 114-656 &   22 41 35 & +01 11 12  & 13.096~~~0.002 & 1.947~~~0.008  & 0.767~~~0.002 & 0.293~~~0.002  & 12.644~~~0.003 & 0.698~~~0.022  & 0.965~~~0.007 & 0.506~~~0.002& 1 \\
SA 114-548 &   22 41 37 & +00 59 07  & 12.290~~~0.003 & 3.111~~~0.009  & 1.147~~~0.004 & 0.441~~~0.003 & 11.601~~~0.001 & 1.573~~~0.006  & 1.362~~~0.002 & 0.651~~~0.001& 1 \\
G 27-45    &   22 44 56 &$-$02 21 13  & 11.783~~~0.002 & 1.113~~~0.004  & 0.502~~~0.003 & 0.193~~~0.003 & 11.500~~~0.005 & -0.080~~~0.010 & 0.680~~~0.007 & 0.400& 1        \\
BD +33 4737 &  23 34 36 & +34 02 22  &  9.415~~~0.002 & 1.650~~~0.004  & 0.576~~~0.003 & 0.177~~~0.003  &  9.040         & 0.390          & 0.800        &    0.330& 1    \\
PG 2336+004B & 23 38 38 & +00 42 47  & 12.636~~~0.003 & 1.113~~~0.007  & 0.328~~~0.003 & 0.101~~~0.002  & 12.431         & -0.035~~~0.002 & 0.507~~~0.002 & 0.316~~~0.001& 1 \\
SA 115-420 &   23 42 36 & +01 05 59  & 11.339~~~0.002 & 1.105~~~0.004  & 0.281~~~0.003 & 0.081~~~0.002  & 11.161         & -0.027~~~0.002 & 0.468~~~0.001 & 0.293~~~0.001& 1 \\ 
\end{tabular}  
}  
\end{table*}
\end{document}